\documentclass{article}
\usepackage{ijcai26}

\usepackage{times}
\usepackage{soul}
\usepackage{url}
\usepackage[hidelinks]{hyperref}
\usepackage[utf8]{inputenc}
\usepackage[small]{caption}
\usepackage{graphicx}
\usepackage{amsmath}
\usepackage{amsthm}
\usepackage{booktabs}
\usepackage{algorithm}
\usepackage{algorithmic}
\usepackage[switch]{lineno}
\usepackage{amsmath}
\usepackage{booktabs}
\usepackage{multirow}
\usepackage{tikz}
\usepackage[most]{tcolorbox}   % 用于显示那个灰色的 Prompt Box
\usepackage{amssymb}   % 修复 \checkmark (对号)
\usepackage{xcolor}    % 修复 \textcolor (红/绿色)
\usepackage{enumitem}    % 用于 Box 里的列表格式调整 
\usepackage{pgf-pie}
\usepackage{microtype}
\usepackage{inconsolata}
\usepackage{graphicx}
\usepackage[space]{grffile}
\usepackage{CJKutf8} % 新加入
\usepackage{url}

\title{Benchmarking Real-Time Question Answering via Executable Code Workflows}

\author{
Wenjie Zhou$^1$\thanks{These authors contributed equally.} \and
Yuan Gao$^2$\footnotemark[1]\thanks{Work done during the internship at Li Auto Inc.} \and
Xin Zhou$^1$\thanks{Corresponding author.} \and
Hao Fu$^1$ \And
Zhongjian Miao$^1$ \and
Wei Chen$^1$ \and
Bo Chen$^2$ \and
Xiaobing Zhao$^2$\\
\affiliations
$^1$Li Auto Inc.\\
$^2$Minzu University of China\\
\emails
\{zhouwenjie1, zhouxin12, miaozhongjian, chenwei10\}@lixiang.com,\\
\{ignitesun, nmzxb\_cn\}@163.com,
haofuch@outlook.com,
chenbomuc@muc.edu.cn
}
\begin{document}
\maketitle

\begin{abstract}

Retrieving real-time information is a fundamental capability for search-integrated agents in real-world applications. However, existing benchmarks are predominantly static and therefore fail to capture the temporal dynamics of information and the continuously evolving nature of real-world knowledge. 
To address this limitation, we propose \textbf{RT-QA}, a dynamic evaluation framework that leverages executable code workflows to retrieve up-to-date answers at evaluation time.
Specifically, we construct an agent-driven pipeline to automatically generate Python workflows for web crawling and DOM-based answer extraction. 
RT-QA spans 12 domains with 320 Chinese questions categorized into three difficulty levels. 
To ensure benchmark reliability, all final problem-workflow pairs undergo workflow execution verification and manual website validation. Extensive evaluations of state-of-the-art models reveal significant limitations in real-time adaptability: even the best-performing models achieve only around 46\% accuracy.
Our analysis highlights two primary failure modes: 
(1) \textbf{Lazy Retrieval}, where agents rely on search snippets instead of deeply scanning specific websites for information; and 
(2) \textbf{Temporal Confusion}, a cognitive error where agents retrieve a historical date (e.g., an event in 2024) and fail to re-anchor to the current time (2026) for subsequent reasoning. 
These findings suggest that future agents require not just better retrieval strategies, but robust temporal state management.
The code and data are available at \url{https://github.com/leaves-slient/RT-Bench.git}.
\end{abstract}

\section{Introduction}

\begin{figure}[t]
    \centering
    \includegraphics[width=\linewidth]{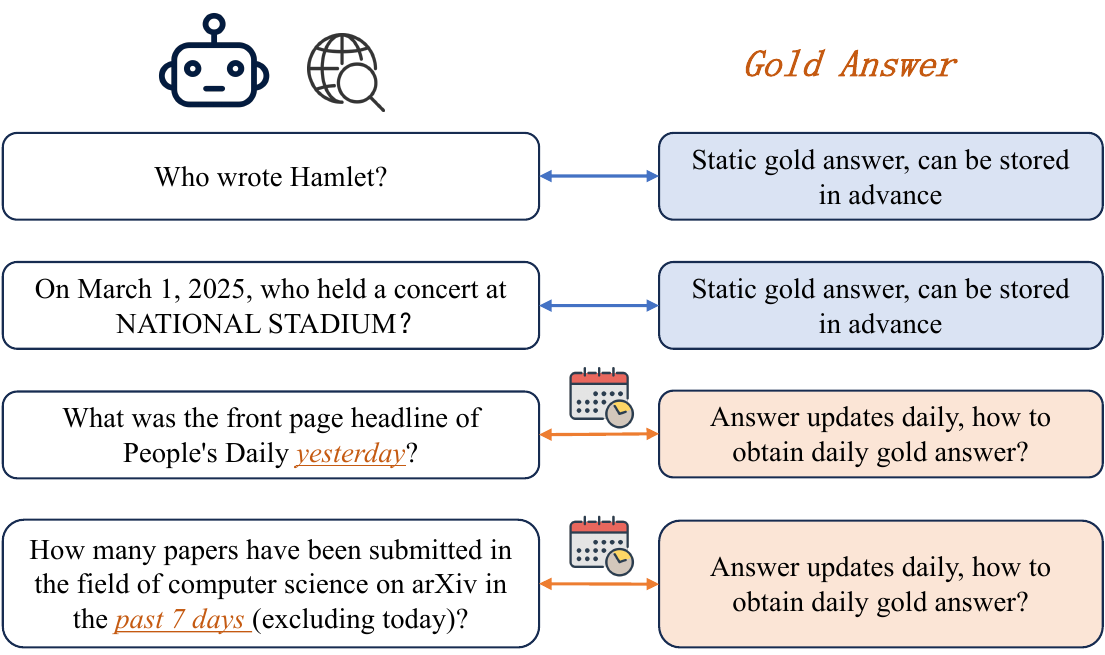} 
    \caption{\textbf{Comparison between Static QA and Real-time QA.} Unlike static questions (top) where answers are fixed or can be pre-stored, real-time questions (bottom) typically involve relative temporal constraints (e.g., \textit{``yesterday''}, \textit{``past 7 days''}).}
    \label{fig:concept}
\end{figure}

The capability to access and reason over real-time information is a fundamental requirement for general intelligence. In recent years, Large Language Models (LLMs) have evolved from static knowledge bases into dynamic, search-integrated agents. The emergence of commercial systems like SearchGPT \cite{openai2024searchgpt}, Perplexity \cite{perplexity2024}, and the DeepResearch framework \cite{team2025deepresearch} highlights this shift, demonstrating impressive performance in complex information-seeking tasks \cite{nakano2021webgpt,schick2024toolformer} and effectively transitioning LLMs into comprehensive assistants.

% Despite this rapid development, evaluation methodologies for such agents remain limited. The majority of existing benchmarks rely on \textbf{static} question-answer pairs \cite{rajpurkar2016squad,kwiatkowski2019natural,yang2023simpleqa}, suffering from two critical flaws.
% First, Data Leakage leads to inflated scores, as models often answer questions by "memorizing" pre-training data rather than performing genuine retrieval.
% Second, Knowledge Obsolescence renders benchmarks ineffective for dynamic queries, as they cannot penalize agents that hallucinate based on outdated facts instead of fetching real-time updates.

Despite this rapid development, evaluation methodologies for such agents remain limited. The majority of existing benchmarks rely on \textbf{static} question-answer pairs \cite{rajpurkar2016squad,kwiatkowski2019natural,yang2023simpleqa}, suffering from two critical flaws: \textbf{Data Leakage} and \textbf{Knowledge Obsolescence}. 
The former enables agents to answer from memory instead of conducting necessary searches, while the latter renders the evaluation ground truth invalid as facts evolve.

To mitigate this, recent studies have introduced periodic benchmarks that update monthly or annually \cite{liska2022streamingqa,vu2023freshllms,kasai2022realtime}. While some works propose automated pipelines \cite{cheng2024pat,cheng2025dailyqa}, they remain limited by their dependency on pre-existing structured sources (e.g., Wikidata or specific APIs).
Ideally, answering daily questions—such as tracking stock prices or arXiv submissions—requires querying official APIs. But such APIs are often scarce, paywalled, or simply do not exist. Consequently, existing benchmarks must compromise, covering only unchanged facts or a few API-supported domains. This leaves a critical gap: agents are rarely tested on high-frequency information from the open web, where no direct API exists.

In this paper, we introduce \textbf{RT-QA}, a dynamic benchmark designed to evaluate agents on rapidly changing, daily-updated information. The core innovation of RT-QA is the paradigm shift from Static Answers to \textbf{Executable Code Workflows}. 
Instead of a fixed answer string, each test item in RT-QA consists of a natural language question paired with a Python-based workflow. This workflow functions by visiting authoritative websites, parsing unstructured DOM elements, and generating the real-time ground truth at the exact moment of evaluation. To realize this, we propose the \textbf{RT Agent} framework, which leverages the tool-creation capabilities of LLMs to automatically generate these workflows and includes a self-repair mechanism to handle inevitable changes in web page structures. Finally we construct high-value questions across 12 domains (including Finance, Public Policy, and Science) that update daily, weekly, or monthly, minimizing manual effort.

We conducted extensive evaluations of state-of-the-art agents, including GPT-5.2, Claude-4.5, and GLM-4.7 \cite{5team2025glm45agenticreasoningcoding}, on RT-QA.
The results reveal a significant gap: while these models excel on static tasks, they struggle with real-time inquiries.
Even the leading models achieve an average accuracy of only 46\%, with many other models falling below 40\%.
To understand the root causes, we conducted a large scale analysis of over 7,000 agent trajectories.
Our taxonomy of failure modes reveals that the primary bottleneck remains \textbf{information gathering}: Retrieval-related issues (including Lazy Retrieval, Incomplete Scan, and Source Quality) account for 45\% of all errors. Specifically, 20\% of failures stem from Lazy Retrieval, where agents rely on search snippets without visiting the actual page.

Moreover, we uncover a recurring reasoning error: Temporal Confusion. We observe that agents frequently perform worse on intermediate reasoning tasks (Level 2) than on highly complex ones (Level 3).
Our analysis suggests this stems from poor temporal state management—agents mistakenly use historical dates retrieved in intermediate steps as the ``current time'' for subsequent reasoning, whereas the higher complexity of L3 tasks forces a more rigorous isolation of temporal variables.

Our contributions are summarized as follows:
\begin{itemize}
    \item We propose RT-QA, a benchmark using executable workflows to dynamically generate ground truth at inference time.
    \item We develop the \textbf{RT Agent}, an automated framework for generating and maintaining these workflows, allowing users to easily extend the benchmark.
    \item We provide a comprehensive evaluation and a fine-grained failure taxonomy. We highlight \textbf{Lazy Retrieval} and \textbf{Temporal Confusion} as the two dominant failure modes, offering critical insights for future agents to improve retrieval depth and temporal state management.
\end{itemize}

% =========================================================
% FIGURE 3: Workflow Code Example (Wide Figure)
% =========================================================
\begin{figure*}[h!]
    \centering
    \includegraphics[width=\linewidth]{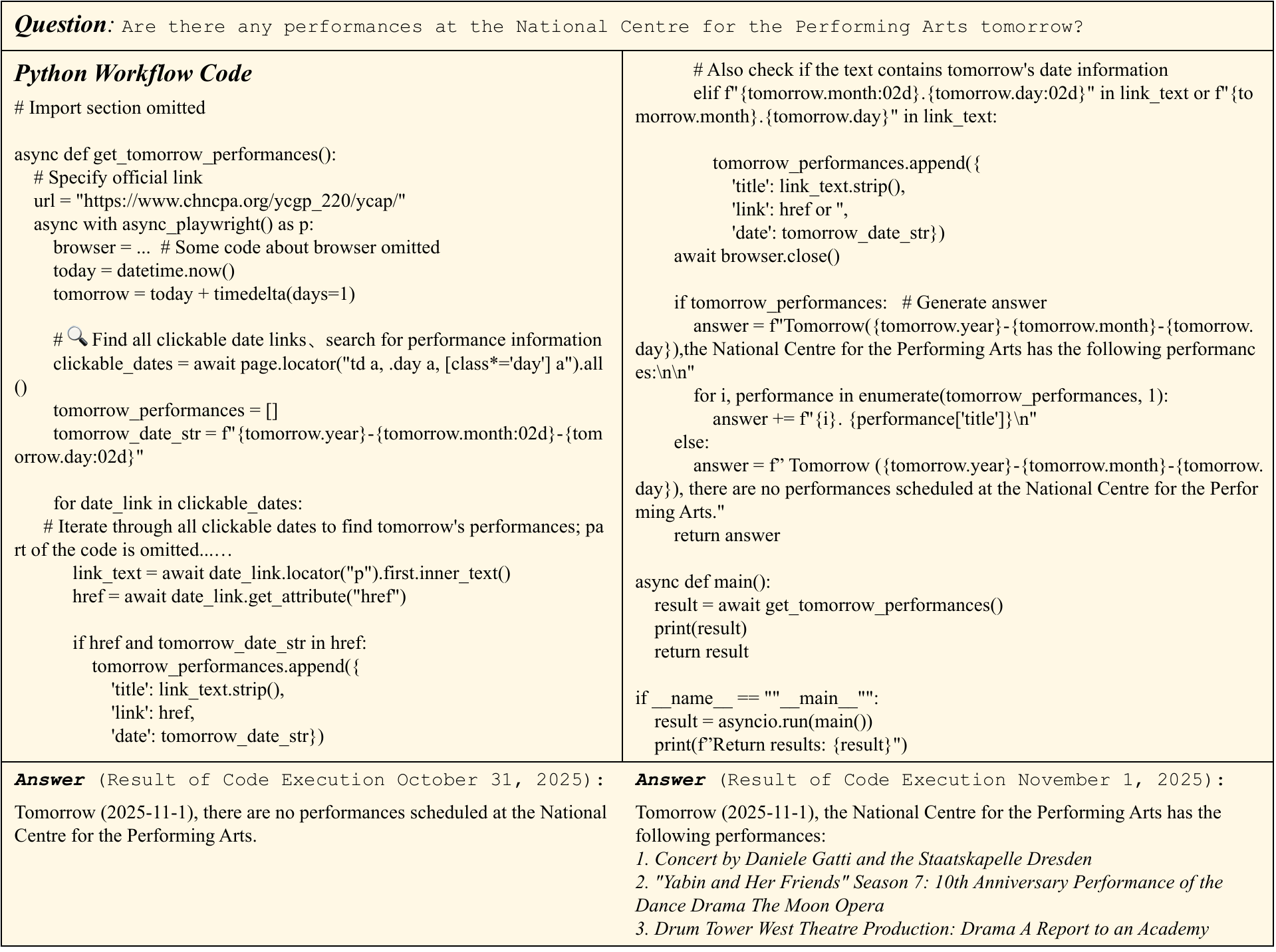} 
    \caption{\textbf{An Example of an RT-QA Executed Workflow.} The figure demonstrates how a single Python workflow handles dynamic time anchoring. The code (left) explicitly calculates the target date (tomorrow) based on the execution time. Consequently, executing the same workflow on different dates (Oct 31 vs. Nov 1) yields correct, context-aware answers (bottom right), functioning as a dynamic pseudo-API.}
    \label{fig:code_example}
\end{figure*}

\section{Related Work}

\subsection{From Static RAG to Real-Time Agents}
While LLMs excel at reasoning, their internal knowledge is static. To mitigate this Knowledge Obsolescence, Retrieval-Augmented Generation (RAG) methods \cite{lewis2020retrieval,izacard2022few} connect models to external corpora. Recent works like TimeRAG \cite{wang2025timerag} and Time-Aware LMs \cite{dhingra2022timeaware} further improved temporal grounding. However, standard RAG relies on pre-indexed documents, which inevitably suffer from indexing latency—they cannot answer questions about events that happened minutes ago.
To overcome this, the field is shifting towards autonomous agents that interact directly with the live web. Early works like WebGPT \cite{nakano2021webgpt} and WebArena \cite{zhou2023webarena} demonstrated how agents can use browsers as tools. More recently, advanced frameworks like DeepResearch \cite{team2025deepresearch} have enabled agents to perform concurrent browsing and complex planning. Our work builds on this trend and focuses on evaluation: providing a rigorous benchmark to measure how well these agents actually perform in dynamic, real-world scenarios.

\subsection{Benchmarks for Time-Sensitive QA}
Evaluating temporal reasoning has evolved significantly. Early datasets like TempQuestions \cite{jia2018tempquestions} and Situational QA \cite{zhang2022situational} focused on static text. Recent dynamic benchmarks have pushed the boundary: StreamingQA \cite{liska2022streamingqa} and FreshLLMs \cite{vu2023freshllms} introduce periodic updates (e.g., monthly) to track world events, while DailyQA \cite{cheng2025dailyqa} leverages Wikipedia revision logs.
However, a fundamental distinction exists between these approaches and RT-QA. Existing benchmarks operate on a Static Answers paradigm: data is collected first and released at fixed intervals. This introduces an unavoidable latency gap—they cannot evaluate queries about events happening today or right now. 
In contrast, RT-QA ensures nearly zero latency. This allows us to cover high-frequency web information (e.g., live stock prices, instant weather warnings) that periodic benchmarks inevitably miss.

\subsection{Agent-Driven Tool Creation}
Recent breakthroughs in LLMs, particularly with models like GPT-5.2 \cite{openai2025gpt5}, Claude 4.5, and DeepSeek-V3.2 \cite{liu2025deepseek}, have redefined autonomous code generation. Beyond merely utilizing existing APIs \cite{schick2024toolformer,patil2023gorilla}, advanced agents can now synthesize complex executable logic to interact with dynamic environments. This trend is exemplified by autonomous software engineers like Devin \cite{cognition2024devin} and OpenDevin \cite{opendevin2024}, as well as generalist web agents like Voyager \cite{wang2023voyager} and Mind2Web \cite{deng2023mind2web}.
RT-QA leverages this tool-creation capability to revolutionize benchmark construction. Instead of relying on manual scripting or rigid templates, we implement an Agent-in-the-Loop pipeline. By harnessing the advanced reasoning and coding capabilities of these models, our RT Agent autonomously analyzes unstructured web pages and generates precise Python workflows. This approach allows us to scale the construction of dynamic benchmarks far beyond what was previously possible with manual effort.

% =========================================================
% FIGURE 2: Automated Construction Pipeline (Wide Figure)
% =========================================================
\begin{figure*}[t]
    \centering
    \includegraphics[width=\linewidth]{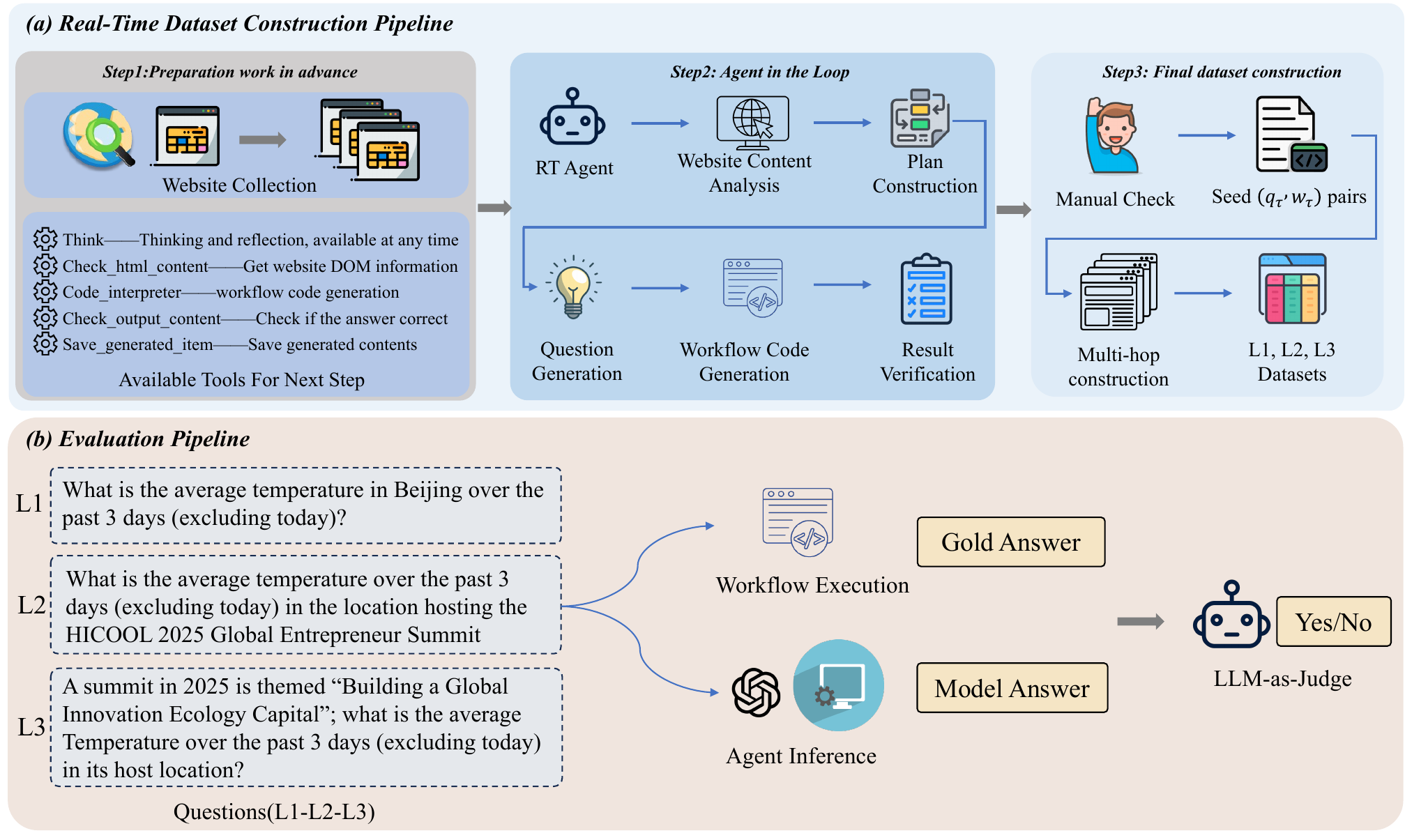} 
    \caption{\textbf{Overview of the RT-QA Framework.} (a) The Construction Pipeline consists of three steps: preparation and tool definition, Agent-in-the-Loop workflow generation, and final dataset construction. (b) The Evaluation Pipeline, where models and workflows are executed simultaneously to assess real-time accuracy.}
    \label{fig:pipeline}
\end{figure*}

\section{RT-QA Benchmark}
\label{sec:method}

RT-QA is a dynamic benchmark designed to evaluate agents on rapidly changing, daily-updated information. 

\subsection{Problem Formulation}
\label{sec:formulation}
Standard QA benchmarks represent a test sample as a static pair $(q, a)$. However, this formulation is insufficient for real-time scenarios where the ground truth is time-dependent. In RT-QA, we redefine a test item as a tuple $(q_\tau, w_\tau)$, where:
\begin{itemize}
    \item $q_\tau$ is a natural language question anchored to a relative time constraint (e.g., ``yesterday'', ``past 7 days'').
    \item $w_\tau$ is an \textbf{Executable Code Workflow} serving as a dynamic retrieval script. It encapsulates the logic to visit a target URL and extract specific information by parsing raw DOM elements.
\end{itemize}

The ground truth answer $a_t$ at any evaluation timestamp $t$ is obtained dynamically:
\begin{equation}
a_t = \text{Exec}(w_\tau, t)
\end{equation}
As illustrated in Figure \ref{fig:code_example}, the execution function $\text{Exec}(\cdot)$ three key steps:
(1) Temporal Resolution: converting relative terms in $q_\tau$ to absolute dates based on $t$;
(2) Live Retrieval: navigating to the target website via a headless browser; and
(3) DOM Parsing: locating and computing the target elements to produce a deterministic answer.

\subsection{Real-Time Dataset Construction Pipeline}
\label{sec:pipeline}
Constructing robust executable workflows needs to process complex, dynamic DOM structures. To automate this, we propose an Agent-in-the-Loop pipeline (Figure \ref{fig:pipeline}a). We implement the RT Agent (powered by Claude 4.5), equipped with the specific toolset listed in Step 1 of the pipeline. The process operates in three specific steps:

\paragraph{Step 1: Preparation Work.}
We curate a diverse set of authoritative websites across 12 domains (e.g., National Bureau of Statistics, Stock Exchanges) that update frequently. Simultaneously, we define the \textit{Available Tools} (e.g., \texttt{Check\_html\_content}, \texttt{Code\_interpreter}) that will be provided to the agent in the subsequent step.

\paragraph{Step 2: Agent-in-the-Loop.}
Given a target URL, the RT Agent autonomously generates both the question and the corresponding workflow through an iterative process:
\begin{itemize}
    \item \textit{Exploration \& Coding:} The agent utilizes \texttt{Check\_html\_content} tool (backed by Playwright) to fetch the raw DOM tree and identify dynamic elements. It then uses the \texttt{Think} tool to plan a strategy and writes Python scripts via the \texttt{Code\_interpreter} tool. We enforce strict coding standards that prohibit silent failure (e.g., broad try-except blocks), ensuring that runtime errors are propagated back to the agent for debugging and self-repair.
    
    \item \textit{Result Verification:} Once a candidate code workflow is generated, the agent invokes \texttt{Check\_output\_content}. This tool executes the generated code to retrieve the text answer while simultaneously capturing a page screenshot. The agent then cross-validates the text against the visual evidence to ensure alignment with human perception.
\end{itemize}

\paragraph{Step 3: Final Dataset Construction.}
The output from Step 2 is a set of candidate pairs $(q_\tau, w_\tau)$. We then apply a rigorous two-phase process to finalize the dataset:
\begin{itemize}
\item \textit{Phase 1: Human Validation.} We conduct a comprehensive manual review on the candidate pairs. Annotators verify that the generated questions ($q_\tau$) are fluent and unambiguous, and that the workflows ($w_\tau$) consistently retrieve the correct information. Verified pairs serve as the L1 (Direct) seed dataset.
\item \textit{Phase 2: Multi-hop Expansion.} To construct L2 and L3 datasets, we manually expand these L1 seeds based on the reasoning logic defined in Table \ref{tab:statistics}. For example, an L1 question about ``Beijing's temperature'' is expanded into an L2 question by replacing ``Beijing'' with an entity description (e.g., ``the host city of the 2025 HICOOL Summit''). By reusing the verified L1 workflow for L2 and L3, we ensure the ground truth remains consistent, regardless of how complex the question becomes.
\end{itemize}

During construction, the RT Agent successfully returned generated workflows for 71\% of attempted cases, and 96.3\% of these returned workflows passed manual validation. All final problem-workflow pairs were manually reviewed before formal evaluation while others were revised or discarded. The resulting benchmark comprises a total of \textbf{320 test samples} spanning 12 domains, categorized into \textbf{L1 (Direct, $N=154$)}, \textbf{L2 (1-Hop, $N=109$)}, and \textbf{L3 (2-Hop+, $N=59$)}. This design tests whether an agent can switch between retrieving historical facts to identify the target entity and fetching real-time data to answer the core question.

\paragraph{Human Effort and Cost.}
As we mentioned before, RT-QA is not fully automatic and final human verification is required, the agent-assisted construction pipeline substantially reduces manual effort. Without the RT Agent, manually designing and maintaining one real-time QA workflow takes approximately 4 hours per case. With the proposed pipeline, the required human time is reduced to about 0.05 hours per case, mainly for final verification. The token cost is approximately 160k tokens per case. 

% Table 1 Code (Kept exactly as you provided, looks good)
\begin{table}[t]
\centering
\small
\renewcommand{\arraystretch}{1.1} 
\resizebox{\linewidth}{!}{
\begin{tabular}{lclc}
\toprule
\multicolumn{4}{c}{\textbf{Panel A: Domain Distribution (12 Categories)}} \\
\midrule
\textbf{Domain} & \textbf{\#} & \textbf{Domain} & \textbf{\#} \\
\midrule
Finance \& Economy & 42 & Public Safety & 34 \\
Weather \& Env. & 40 & News \& Media & 33 \\
Culture Activities & 35 & Space \& Astronomy & 28 \\
Sports Events & 28 & Energy \& Industry & 20 \\
Consumer Products & 20 & Transportation & 18 \\
Sci \& Tech & 14 & Policy \& Gov. & 10 \\
\midrule
\end{tabular}
}
\resizebox{\linewidth}{!}{
\begin{tabular}{p{0.8cm} p{1.2cm} p{6.5cm}} 
\multicolumn{3}{c}{\textbf{Panel B: Reasoning Complexity \& Logic}} \\
\midrule
\textbf{Level} & \textbf{Hops} & \textbf{Example Question (Translated) \& Logic} \\
\midrule
\textbf{L1} & \textbf{0-Hop} & ``What is the average temperature in \textbf{Beijing} over the past 3 days (excluding today)?'' 
\textit{\textcolor{gray}{$\hookrightarrow$ Logic: Locate Beijing $\rightarrow$ Search.}} \\ 
\addlinespace[8pt]

\textbf{L2} & \textbf{1-Hop} & ``What is the avg. temp. over the past 3 days (excluding today) in the location hosting the \textbf{HICOOL 2025 Global Entrepreneur Summit}?''  
\textit{\textcolor{gray}{$\hookrightarrow$ Logic: HICOOL 2025 Summit $\xrightarrow{Map}$ Beijing $\rightarrow$ Search.}} \\ 
\addlinespace[8pt]

\textbf{L3} & \textbf{2-Hop+} & ``A summit in 2025 is themed \textbf{`Building a Global Innovation Ecology Capital'}; what is the avg. temp. over the past 3 days (excluding today) in its host location?''  
\textit{\textcolor{gray}{$\hookrightarrow$ Logic: Theme $\xrightarrow{Map}$ HICOOL 2025 $\xrightarrow{Map}$ Beijing $\rightarrow$ Search.}} \\
\bottomrule
\end{tabular}
}
\caption{\textbf{RT-QA Statistics.} Top: Domain distribution. Bottom: Difficulty definitions based on reasoning hops. \textit{Note: Example questions are translated from Chinese to demonstrate the reasoning complexity.}}
\label{tab:statistics}
\end{table}

\subsection{Maintenance \& Extensibility}

\paragraph{Automated Self-Repair.}
Although all workflows are verified before inclusion, website layouts may change over time, causing previously valid selectors or extraction logic to fail. To ensure long-term stability, we deploy a Repair Agent with a similar architecture to the RT Agent. When a workflow fails during execution, for example by throwing an exception, returning an empty result, or producing an obviously invalid answer, the failed case is sent to the Repair Agent. The Repair Agent re-analyzes the live DOM of the target website, identifies structural changes such as modified class names or shifted content blocks, and patches the Python workflow accordingly.

During the formal evaluation period, about 5\% of workflows required repair due to website-side changes. For the observed repair cases, the Repair Agent successfully restored execution in all cases after re-analyzing the live page and updating the extraction logic. This maintenance burden is acceptable in our process. 

\paragraph{Extensibility as a Service.}
Beyond the initial 320 questions, RT-QA serves as an open framework. Researchers can simply provide a target URL and a topic of interest; our system will automatically synthesize questions and verify the corresponding workflows. This allows the community to easily expand RT-QA into specific domains without manual coding.

\section{Experiments}
\label{sec:experiments}

\subsection{Experimental Setup}

\begin{table*}[t]
\centering
\small
\renewcommand{\arraystretch}{1.25} % 稍微增加行高，让数据不拥挤
\setlength{\tabcolsep}{4pt}        % 稍微收紧列间距，防止表格太宽
\caption{\textbf{Main Results on RT-QA (Jan 10--15, 2026).} The table reports the \textbf{Accuracy (\%)} evaluated by GPT-5.2 Judge. The left section shows the daily trend. The right section breaks down average performance by different level. \textbf{Total Avg} represents the aggregated accuracy across the entire 6-day period for all questions.}
\label{tab:main_results}

\begin{tabular}{l cccccc ccc c}
\toprule
% 第一行表头
\multirow{2}{*}{\textbf{Model}} & \multicolumn{6}{c}{\textbf{Daily Overall Trend (Timeline)}} & \multicolumn{3}{c}{\textbf{Level Performance}} & \multirow{2}{*}{\textbf{Total Avg}} \\
% 横线分组
\cmidrule(lr){2-7} \cmidrule(lr){8-10}
% 第二行表头
 & \textbf{Jan 10} & \textbf{Jan 11} & \textbf{Jan 12} & \textbf{Jan 13} & \textbf{Jan 14} & \textbf{Jan 15} & \textbf{L1} & \textbf{L2} & \textbf{L3} & \\
\midrule
\multicolumn{11}{l}{\textit{\textbf{Open-Weights Models}}} \\
GLM-4.7 & 43.6 & 46.0 & 46.1 & 48.4 & 43.8 & 43.1 & \textbf{46.9} & 41.9 & 46.9 & \textbf{45.2} \\
DeepSeek-V3.2 & 30.1 & 41.6 & 33.9 & 34.4 & 34.2 & 36.3 & 34.2 & 34.0 & \textbf{39.3} & 35.1 \\
Qwen3-235B & 24.5 & 29.9 & 34.2 & 25.6 & 34.5 & 35.6 & \textbf{33.0} & 26.9 & 31.9 & 30.7 \\
Kimi-K2 & 37.6 & 41.5 & 40.4 & 40.6 & 41.0 & 41.6 & \textbf{43.0} & 37.4 & 39.6 & 40.4 \\
\midrule
\multicolumn{11}{l}{\textit{\textbf{Proprietary Models}}} \\
Doubao-Seed-1.8 & 46.9 & 47.4 & 49.4 & 45.6 & 43.2 & 46.3 & 44.3 & 45.6 & \textbf{53.7} & \textbf{46.5} \\
GPT-5.2 & 38.2 & 42.2 & 44.7 & 38.4 & 36.6 & 45.9 & \textbf{43.2} & 38.0 & 41.0 & 41.0 \\
Gemini-3-Pro & 34.8 & 36.6 & 38.1 & 37.0 & 39.1 & 38.4 & \textbf{41.0} & 33.2 & 35.4 & 37.3 \\
Claude-4.5 & 42.9 & 43.4 & 48.1 & 42.2 & 40.7 & 40.9 & 44.3 & 39.2 & \textbf{46.9} & 43.0 \\
\bottomrule
\end{tabular}
\end{table*}

\paragraph{Agent Framework for evaluation.}
Evaluating agents via official web interfaces (e.g., ChatGPT-Web) presents two challenges: lack of transparency (intermediate steps are invisible) and scalability issues (manual testing is labor-intensive). Therefore, to ensure reproducibility and fine-grained analysis, we standardize our evaluation using an controllable open-source agent framework (adapted from DeepResearch \cite{team2025deepresearch}) powered by the respective model APIs.To tailor the framework for real-time QA, we introduced two key modifications compared to the original implementation:

(1) Native Function Calling \& Toolset: The original framework relies on XML-based prompting for tool use. We replaced this with native function calling interfaces to enhance compatibility and robustness. Under this protocol, agents are equipped with two main tools:
\begin{itemize}[leftmargin=*, topsep=2pt, itemsep=2pt]
\item \texttt{Search}: Powered by Serper API, retrieving top-10 results. It includes an automatic locale routing strategy (e.g., mapping Chinese queries to \texttt{cn}) to ensure linguistic relevance.
\item \texttt{Visit}: Uses the Jina Reader API to fetch page content. To handle context limits, we implement a Goal-Driven Extraction mechanism where the model filters raw content to extract only evidence relevant to the current browsing goal.
\end{itemize}

(2) Temporal Injection: We inject the precise current timestamp into the system prompt. This enforces a "Real-Time Protocol," instructing agents to convert relative expressions (e.g., \textit{yesterday}) into absolute dates before tool execution. The full prompt for model generation is provided in Appendix~\ref{app:model_generation_prompt}.

\paragraph{Prompt Optimization}
To examine whether performance can be substantially improved by prompt engineering alone, we use Qwen3-235B as a representative backbone and keep the agent framework, tools, decoding setting, and judge unchanged while varying only the system prompt. Starting from the original DeepResearch prompt, accuracy improves from 16.0\% to 24.6\% after emphasizing real-time awareness, further to 28.2\% after requiring time anchoring and relative-time resolution, and finally to 30.6\% after explicitly distinguishing retrieved historical dates from the current evaluation time. These gains confirm the importance of temporal instructions, but the diminishing returns suggest that the remaining bottleneck lies beyond prompting and requires deeper retrieval, stronger evidence verification, and explicit temporal state management.

\paragraph{Evaluation Setup.}
We evaluated 8 state-of-the-art models (as of Jan 2026), including proprietary ones (GPT-5.2, Gemini-3-Pro, Claude-4.5, Doubao-Seed) and open-weights models (DeepSeek-V3.2, GLM-4.7, Qwen3, Kimi-K2), all set to temperature 
temperature $T=0.6$.
Experiments were conducted over a 6-day period from January 10 to January 15, 2026. 
As shown in the Evaluation Pipeline (Figure \ref{fig:pipeline}b), for each question, the ground truth $a_t$ is generated by executing the workflow $w_\tau$ at the exact same timestamp as the model inference. 
To assess correctness, we adopt the LLM-as-a-Judge paradigm \cite{zheng2023judging}. 
We utilize GPT-5.2 as the evaluator to assess semantic equivalence between the model's response and the dynamic ground truth. 
\textit{(Note: GPT-5.2 was selected due to its high correlation with human annotations in our validation. In a human agreement check on Qwen3-235B's responses to 109 questions, three independent GPT-5.2 judging runs agreed with human annotations on 109/109, 109/109, and 108/109 cases, respectively.)}

\subsection{Main Results}
Table \ref{tab:main_results} presents the comprehensive performance of 8 leading models over a continuous 6-day evaluation period (Jan 10--15, 2026).
\paragraph{Low Overall Accuracy in Real-Time Settings.}
Contrary to their high performance on static benchmarks, models struggle significantly with real-time information. Even the leading models (Doubao-Seed and GLM-4.7) achieve an average accuracy of only 46\%, while most open-weights models fall below 40\%. This confirms that current agents still lack reliable information-seeking capabilities for the open web.
\paragraph{High Volatility Caused by Lazy Retrieval.}
Model performance fluctuates drastically day-to-day. For instance, DeepSeek-V3.2 varied by over 11 points (30.1\% to 41.6\%) within 24 hours. Our analysis links this instability to \textit{Lazy Retrieval}. Agents often rely on search snippets rather than navigating to authoritative websites for step-by-step verification. Consequently, their success depends largely on whether the search engine surfaces a direct answer or relevant article in its immediate results, leading to inconsistent results.

This underscores the critical importance of \textit{genuine deep search}, where agents must actively browse and verify source content instead of passively consuming summaries.

\begin{figure}[h]
    \centering
    \includegraphics[width=\columnwidth]{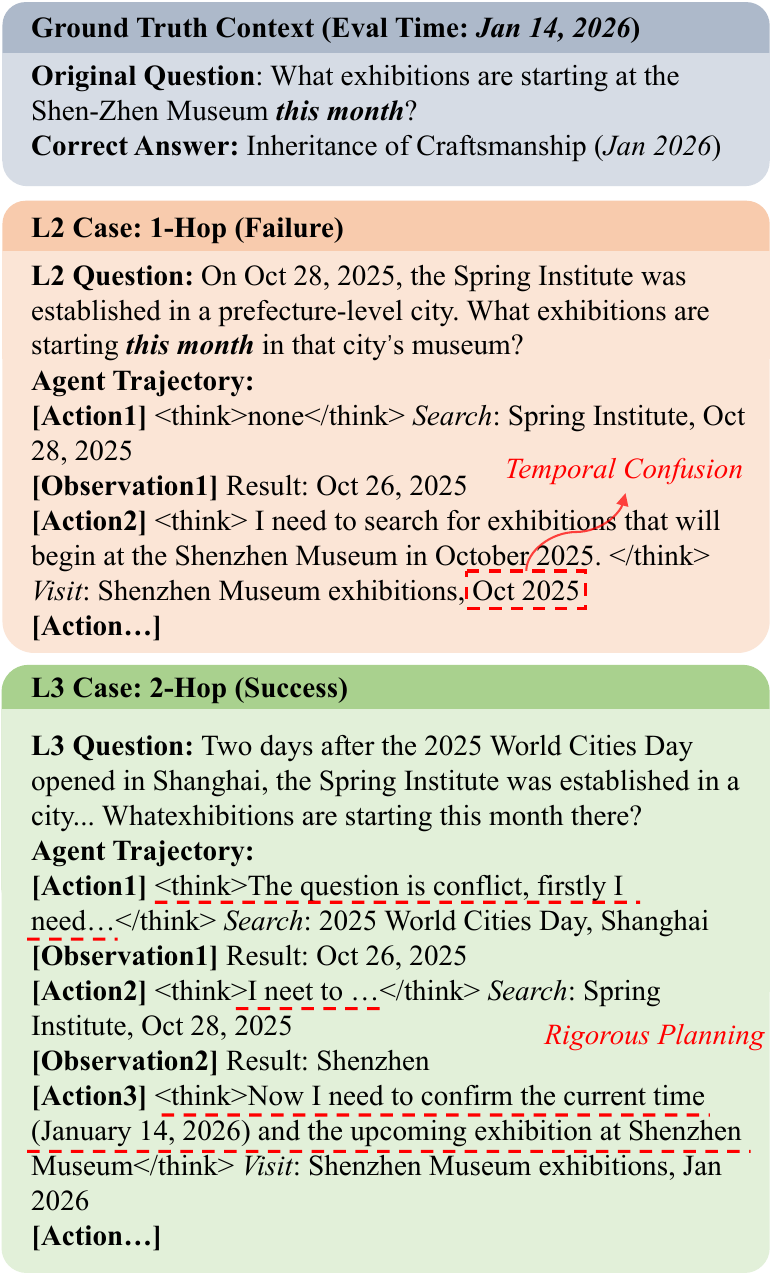} 
    \caption{\textbf{Case Study: Temporal Confusion (L2) vs. Rigorous Planning (L3).} 
    Comparison of agent trajectories on the Shenzhen Museum question.}
    \label{fig:case_shenzhen}
\end{figure}

\paragraph{Harder Tasks Perform Better.}
We initially expected performance to degrade linearly with difficulty ($L1 > L2 > L3$). However, the results exhibit a \textbf{U-shaped pattern}, where intermediate-level (L2) tasks often yield the lowest accuracy. For instance, GPT-5.2 achieves 41.0\% accuracy on the most complex L3 tasks, yet drops to 38.0\% on the ostensibly simpler L2 tasks. A typical example is shown in Figure \ref{fig:case_shenzhen}. 

\begin{itemize}
\item \textbf{L2 Failure (Temporal Confusion):} L2 tasks typically consist of two major phases: first, \textbf{identifying the target entity} associated with a past event, and second, \textbf{retrieving information} regarding this entity's status at the current query time. The first phase introduces a strong historical date (e.g., ``Oct 2025''). We observe that agents often get ``stuck'' on this retrieved date, failing to shift their focus back to the current date (``Jan 2026'') when addressing the second phase.
\item \textbf{L3 Recovery (Rigorous Planning):} L3 tasks require a more intricate reasoning chain involving multiple entities and timeframes. This complexity forces the agent to engage in more rigorous planning, explicitly distinguishing between historical event times and the current evaluation time within its working memory.
\end{itemize}

This contrast highlights that \textit{planning capability} is decisive—complexity can sometimes force a model to plan better, acting as a safeguard against simple retrieval errors.

To verify that L2 failures stem from a lack of explicit temporal planning (rather than inherent difficulty), we conducted an additional experiment. We injected a strict Time Anchor instruction into the system prompt, explicitly warning agents to prioritize the current timestamp over retrieved historical dates.
As shown in Table \ref{tab:l2_l3_comparison}, the results are illuminating. With the Time Anchor, L2 accuracy rises to 40.4\%, surpassing L3 (37.3\%). This reversal confirms our hypothesis: L2 tasks are not intrinsically harder than L3; rather, they are cognitively deceptive. Without explicit guidance, agents fail to plan their temporal alignment, but once alerted (via prompt injection) or forced to strategize (via L3 complexity), they can navigate the timeline correctly.

% \subsection{Prompt Sensitivity}

% We further examine how much agent performance can be improved by prompt engineering alone, using Qwen3-235B as a representative backbone model. We keep the search-agent framework, tools, decoding setting, and judge unchanged, and only vary the system prompt. Starting from the original DeepResearch prompt, the accuracy improves from 16.0\% to 24.6\% when emphasizing the real-time nature of the task, further to 28.2\% when requiring time anchoring and relative-time resolution, and finally to 30.6\% when explicitly distinguishing retrieved historical dates from the current evaluation time. The improvement confirms the importance of temporal instructions, but the diminishing gains after this stage suggest that the remaining bottleneck lies beyond prompt engineering and requires deeper retrieval, stronger evidence verification, and explicit temporal state management.

% \begin{figure}[t]
%     \centering
%     \includegraphics[width=\columnwidth]{images/contract.pdf} 
%     \caption{\textbf{Case Study: Temporal Confusion (L2) vs. Rigorous Planning (L3).} 
%     Comparison of agent trajectories on the Shenzhen Museum question.}
%     \label{fig:case_shenzhen}
% \end{figure}

\subsection{Failure Mode Analysis}
\label{sec:analysis}

To better understand the root causes of low performance, we conducted a large scale analysis of over 7000 error trajectories. As illustrated in \textbf{Figure \ref{fig:error_pie}}, we identify seven distinct failure modes ranging from retrieval strategy to reasoning logic.

\begin{table}[t!]
\centering
\small
\renewcommand{\arraystretch}{1}
\setlength{\tabcolsep}{8pt} % 稍微调整间距以适应长标题
\caption{Performance of L2 and L3 tasks, with a Time Anchor added to the prompt for the Jan 20 setting.}
\vspace{-0.4em}
\label{tab:l2_l3_comparison}

\begin{tabular}{l c c c}
\toprule
% 单行表头直接写
\textbf{Level} & \textbf{Jan 10} & \textbf{Jan 15} & \textbf{Jan 20 (w/ Time Anchor)} \\
\midrule
\textbf{L2} & 33.0 & 40.2 & 40.4 \\
\textbf{L3} & 39.0 & 44.1 & 37.3 \\
\bottomrule
\end{tabular}
\end{table}

% --- Figure 4: Pie Chart (Unchanged) ---
\begin{figure}[h]
    \centering
    \begin{tikzpicture}[scale=0.85] 
        \def\R{2.2} 
        \definecolor{c1}{RGB}{231,76,60}   % Lazy Retrieval. (Red)
        \definecolor{c2}{RGB}{243,156,18}  % Source Reliability (Orange)
        \definecolor{c3}{RGB}{241,196,15}  % Incomplete Scan (Yellow)
        \definecolor{c4}{RGB}{46,204,113}  % Parsing Fail (Green)
        \definecolor{c5}{RGB}{52,152,219}  % Temporal (Blue)
        \definecolor{c6}{RGB}{155,89,182}  % Hallucination (Purple)
        \definecolor{c7}{RGB}{149,165,166} % Logic Error (Gray)

        % 1. Lazy Retrieval. (20%) -> 72 deg
        \draw[fill=c1, draw=white] (0,0) -- (90:\R) arc (90:162:\R) -- cycle;
        \node[font=\tiny, align=center] at (126:\R+0.6) {\textbf{Lazy Retrieval}\\\textbf{20\%}};
        % 2. Source Reliability (10%) -> 36 deg
        \draw[fill=c2, draw=white] (0,0) -- (162:\R) arc (162:198:\R) -- cycle;
        \node[font=\tiny, align=center] at (180:\R+0.8) {\textbf{Source Quality}\\\textbf{10\%}};
        % 3. Incomplete Scanning (15%) -> 54 deg
        \draw[fill=c3, draw=white] (0,0) -- (198:\R) arc (198:252:\R) -- cycle;
        \node[font=\tiny, align=center] at (225:\R+0.7) {\textbf{Incomplete Scan}\\\textbf{15\%}};
        % 4. Parsing Failure (10%) -> 36 deg
        \draw[fill=c4, draw=white] (0,0) -- (252:\R) arc (252:288:\R) -- cycle;
        \node[font=\tiny, align=center] at (270:\R+0.4) {\textbf{Parsing Fail}\\\textbf{10\%}};
        % 5. Temporal (20%) -> 72 deg
        \draw[fill=c5, draw=white] (0,0) -- (288:\R) arc (288:360:\R) -- cycle;
        \node[font=\tiny, align=center] at (324:\R+0.9) {\textbf{Temporal Confusion}\\\textbf{20\%}};
        % 6. Hallucination (15%) -> 54 deg
        \draw[fill=c6, draw=white] (0,0) -- (0:\R) arc (0:54:\R) -- cycle;
        \node[font=\tiny, align=center] at (30:\R+0.7) {\textbf{Hallucination}\\\textbf{15\%}};
        % 7. Logic Error (10%) -> 36 deg
        \draw[fill=c7, draw=white] (0,0) -- (54:\R) arc (54:90:\R) -- cycle;
        \node[font=\tiny, align=center] at (72:\R+0.4) {\textbf{Logic}\\\textbf{10\%}};
    \end{tikzpicture}
    \caption{\textbf{Fine-grained Distribution of Failure Modes.} Failures are decomposed into 7 specific behavioral flaws.}
    \label{fig:error_pie}
\end{figure}

\vspace{-0.8em}

\paragraph{Retrieval \& Scanning Issues (45\%).}
The most prevalent failures stem from inadequate information gathering. Lazy Retrieval (20\%) occurs when agents rely on search snippets without clicking the authoritative websites. Even when visiting the correct page, agents exhibit Incomplete Scanning (15\%), often reading only the first few items of a list. Furthermore, Source Quality (10\%) issues arise when agents prioritize blogs over official government portals.

\paragraph{Temporal Misalignment (20\%).}
Agents struggle with dynamic time anchoring. A common failure is Rigid Grounding, where agents search for a literal date string but fail to recognize data published under relative terms. 
\textbf{Crucially, this category also includes the Temporal Confusion failures observed in L2 tasks, where agents fail to disengage from a retrieved past date (e.g., the event time) and correctly re-anchor to the evaluation time.}

\paragraph{Hallucination \& Logic (25\%).}
When retrieval fails, agents often resort to Hallucination (15\%), Logic Errors (10\%) include calculation mistakes or misclassification.

\section{Conclusion}
\label{sec:conclusion}
In this paper, we introduced \textbf{RT-QA}, a dynamic benchmark designed to evaluate how agents handle high-frequency, real-time information. Instead of relying on static gold answers, RT-QA uses executable workflows to retrieve and compute ground truth at evaluation time. 
Our extensive evaluation reveals that while current agents excel at static reasoning, they remain fragile in dynamic environments.
The discovery of Temporal Confusion reveals a weakness that agents may get stuck in the past, mistakenly using historical dates to answer questions about the present.
This implies that simply giving agents better search tools is not enough. Instead, they need better planning capabilities to distinguish between past events and current goals.
Ultimately, to handle the changing web, agents must evolve from simple searchers into proactive investigators that can actively verify and manage information.
We release RT-QA not just as a dataset, but as an \textbf{extensible infrastructure}.
By democratizing the creation of live benchmarks, we hope to accelerate the transition of Large Language Models from knowledgeable chatbots to truly reliable, real-time assistants.

% \section{Acknowledges}

\bibliographystyle{named}
\bibliography{reference}

@misc{5team2025glm45agenticreasoningcoding,
      title={GLM-4.5: Agentic, Reasoning, and Coding (ARC) Foundation Models}, 
      author={GLM Team and Aohan Zeng and Xin Lv and Qinkai Zheng and Zhenyu Hou and Bin Chen and Chengxing Xie and Cunxiang Wang and Da Yin and Hao Zeng and Jiajie Zhang and Kedong Wang and Lucen Zhong and Mingdao Liu and Rui Lu and Shulin Cao and Xiaohan Zhang and Xuancheng Huang and Yao Wei and Yean Cheng and Yifan An and Yilin Niu and Yuanhao Wen and Yushi Bai and Zhengxiao Du and Zihan Wang and Zilin Zhu and Bohan Zhang and Bosi Wen and Bowen Wu and Bowen Xu and Can Huang and Casey Zhao and Changpeng Cai and Chao Yu and Chen Li and Chendi Ge and Chenghua Huang and Chenhui Zhang and Chenxi Xu and Chenzheng Zhu and Chuang Li and Congfeng Yin and Daoyan Lin and Dayong Yang and Dazhi Jiang and Ding Ai and Erle Zhu and Fei Wang and Gengzheng Pan and Guo Wang and Hailong Sun and Haitao Li and Haiyang Li and Haiyi Hu and Hanyu Zhang and Hao Peng and Hao Tai and Haoke Zhang and Haoran Wang and Haoyu Yang and He Liu and He Zhao and Hongwei Liu and Hongxi Yan and Huan Liu and Huilong Chen and Ji Li and Jiajing Zhao and Jiamin Ren and Jian Jiao and Jiani Zhao and Jianyang Yan and Jiaqi Wang and Jiayi Gui and Jiayue Zhao and Jie Liu and Jijie Li and Jing Li and Jing Lu and Jingsen Wang and Jingwei Yuan and Jingxuan Li and Jingzhao Du and Jinhua Du and Jinxin Liu and Junkai Zhi and Junli Gao and Ke Wang and Lekang Yang and Liang Xu and Lin Fan and Lindong Wu and Lintao Ding and Lu Wang and Man Zhang and Minghao Li and Minghuan Xu and Mingming Zhao and Mingshu Zhai and Pengfan Du and Qian Dong and Shangde Lei and Shangqing Tu and Shangtong Yang and Shaoyou Lu and Shijie Li and Shuang Li and Shuang-Li and Shuxun Yang and Sibo Yi and Tianshu Yu and Wei Tian and Weihan Wang and Wenbo Yu and Weng Lam Tam and Wenjie Liang and Wentao Liu and Xiao Wang and Xiaohan Jia and Xiaotao Gu and Xiaoying Ling and Xin Wang and Xing Fan and Xingru Pan and Xinyuan Zhang and Xinze Zhang and Xiuqing Fu and Xunkai Zhang and Yabo Xu and Yandong Wu and Yida Lu and Yidong Wang and Yilin Zhou and Yiming Pan and Ying Zhang and Yingli Wang and Yingru Li and Yinpei Su and Yipeng Geng and Yitong Zhu and Yongkun Yang and Yuhang Li and Yuhao Wu and Yujiang Li and Yunan Liu and Yunqing Wang and Yuntao Li and Yuxuan Zhang and Zezhen Liu and Zhen Yang and Zhengda Zhou and Zhongpei Qiao and Zhuoer Feng and Zhuorui Liu and Zichen Zhang and Zihan Wang and Zijun Yao and Zikang Wang and Ziqiang Liu and Ziwei Chai and Zixuan Li and Zuodong Zhao and Wenguang Chen and Jidong Zhai and Bin Xu and Minlie Huang and Hongning Wang and Juanzi Li and Yuxiao Dong and Jie Tang},
      year={2025},
      eprint={2508.06471},
      archivePrefix={arXiv},
      primaryClass={cs.CL},
      url={https://arxiv.org/abs/2508.06471}, 
}

@misc{cognition2024devin,
  title = {Devin: The First AI Software Engineer},
  author = {Cognition AI},
  year = {2024},
  howpublished = {\url{https://www.cognition-labs.com/blog}},
  note = {Accessed: 2026-01-15}
}

@article{opendevin2024,
  title={OpenDevin: An Open Platform for AI Software Engineers as Collaborators},
  author={OpenDevin Team},
  journal={arXiv preprint arXiv:2407.16741},
  year={2024}
}

@misc{openai2024searchgpt,
  title = {SearchGPT Prototype: A temporary prototype of new AI search features},
  author = {OpenAI},
  year = {2024},
  howpublished = {\url{https://openai.com/index/searchgpt/}},
  note = {Accessed: 2026-01-15}
}

@misc{perplexity2024,
  title = {Perplexity AI: An AI-powered answer engine},
  author = {Perplexity},
  year = {2024},
  howpublished = {\url{https://www.perplexity.ai}},
  note = {Accessed: 2026-01-15}
}

@article{team2025deepresearch,
  title = {DeepResearch: A Unified Framework for Autonomous Web Navigation and Reasoning},
  author = {Alibaba-NLP Team},
  journal = {Open Source Software},
  year = {2025}
}

@article{zheng2023judging,
  title={Judging LLM-as-a-Judge with MT-Bench and Chatbot Arena},
  author={Zheng, Lianmin and others},
  journal={NeurIPS},
  year={2023}
}

@article{dhingra2022timeaware,
  title={Time-Aware Language Models as Temporal Knowledge Bases},
  author={Dhingra, Bhuwan and others},
  journal={TACL},
  volume={10},
  pages={257--273},
  year={2022}
}

@article{zhou2023webarena,
  title={WebArena: A Realistic Web Environment for Building Autonomous Agents},
  author={Zhou, Shuyan and others},
  journal={ICLR},
  year={2024}
}

@article{wang2023voyager,
  title={Voyager: An Open-Ended Embodied Agent with Large Language Models},
  author={Wang, Guanzhi and others},
  journal={arXiv preprint arXiv:2305.16291},
  year={2023}
}

@article{patil2023gorilla,
  title={Gorilla: Large Language Model Connected with Massive APIs},
  author={Patil, Shishir G and others},
  journal={arXiv preprint arXiv:2305.15334},
  year={2023}
}

@article{zhang2022situational,
    title={Situational QA},
    author={Zhang, Michael J and others},
    journal={EMNLP},
    year={2022}
}

@article{yang2023simpleqa,
  title={Chinese SimpleQA: A Chinese Factuality Evaluation for Large Language Models},
  author={Yang, Yifan and others},
  journal={arXiv preprint arXiv:230x.xxxxx},
  note={Please verify exact arXiv ID based on the PDF description},
  year={2023}
}

@article{openai2025gpt5,
  title={GPT-5 Technical Report},
  author={OpenAI},
  journal={arXiv preprint arXiv:2511.xxxxx},
  year={2025}
}

@article{liu2025deepseek,
  title={Deepseek-v3. 2: Pushing the frontier of open large language models},
  author={Liu, Aixin and Mei, Aoxue and Lin, Bangcai and Xue, Bing and Wang, Bingxuan and Xu, Bingzheng and Wu, Bochao and Zhang, Bowei and Lin, Chaofan and Dong, Chen and others},
  journal={arXiv preprint arXiv:2512.02556},
  year={2025}
}

@inproceedings{cheng2024pat,
  title={PAT-Questions: A Self-Updating Benchmark for Present-Anchored Temporal Question-Answering},
  author={Cheng, Reagan and others},
  booktitle={Proceedings of the 62nd Annual Meeting of the Association for Computational Linguistics (ACL)},
  year={2024}
}

@inproceedings{deng2023mind2web,
  title={Mind2Web: Towards a Generalist Agent for the Web},
  author={Deng, Xiang and others},
  booktitle={NeurIPS},
  year={2023}
}

@inproceedings{lewis2020retrieval,
  title={Retrieval-Augmented Generation for Knowledge-Intensive NLP Tasks},
  author={Lewis, Patrick and Perez, Ethan and Piktus, Aleksandra and others},
  booktitle={Advances in Neural Information Processing Systems (NeurIPS)},
  volume={33},
  pages={9459--9474},
  year={2020}
}

@article{izacard2022few,
  title={Few-shot Learning with Retrieval Augmented Language Models},
  author={Izacard, Gautier and Lewis, Patrick and Lomeli, Maria and others},
  journal={Journal of Machine Learning Research (JMLR)},
  volume={23},
  number={1},
  pages={1--40},
  year={2022}
}

@article{cheng2025dailyqa,
  title={DailyQA: A Benchmark to Evaluate Web Retrieval Augmented LLMs Based on Capturing Real-World Changes},
  author={Cheng, Jiehan and Dou, Zhicheng},
  journal={arXiv preprint arXiv:2505.17162},
  year={2025}
}

@article{wang2025timerag,
  title={TimeRAG: Enhancing Complex Temporal Reasoning with Search Engine Augmentation},
  author={Wang, Zhao and others},
  journal={arXiv preprint arXiv:2511.xxxx}, 
  note={ResearchGate: 3835xxxxx}, 
  year={2025}
}

@article{rajpurkar2016squad,
  title={SQuAD: 100,000+ Questions for Machine Comprehension of Text},
  author={Rajpurkar, Pranav and Zhang, Jian and Lopyrev, Konstantin and Liang, Percy},
  journal={arXiv preprint arXiv:1606.05250},
  year={2016}
}

@article{kwiatkowski2019natural,
  title={Natural Questions: A Benchmark for Question Answering Research},
  author={Kwiatkowski, Tom and Palomaki, Jennimaria and Redfield, Olivia and Collins, Michael and Parikh, Ankur P and Alberti, Chris and Epstein, Danielle and Polosukhin, Illia and Devlin, Jacob and Lee, Kenton and others},
  journal={Transactions of the Association for Computational Linguistics},
  volume={7},
  pages={453--466},
  year={2019}
}

@article{nakano2021webgpt,
  title={WebGPT: Browser-assisted question-answering with human feedback},
  author={Nakano, Reiichiro and Hilton, Jacob and Balaji, Suchir and Wu, Jeff and Ouyang, Long and Kim, Christina and Hesse, Christopher and Jain, Shantanu and Kosaraju, Vineet and Saunders, William and others},
  journal={arXiv preprint arXiv:2112.09332},
  year={2021}
}

@article{schick2024toolformer,
  title={Toolformer: Language models can teach themselves to use tools},
  author={Schick, Timo and Dwivedi-Yu, Jane and Dess{\`\i}, Roberto and Raileanu, Roberta and Lomeli, Maria and Zettlemoyer, Luke and Cancedda, Nicola and Scialom, Thomas},
  journal={Advances in Neural Information Processing Systems},
  volume={36},
  year={2024}
}

@inproceedings{jia2018tempquestions,
  title={TempQuestions: A Benchmark for Temporal Question Answering},
  author={Jia, Zhen and Abujabal, Abdalghani and Saha Roy, Rishiraj and Str{\"o}tgen, Jannik and Weikum, Gerhard},
  booktitle={Proceedings of the 2018 World Wide Web Conference},
  pages={1057--1062},
  year={2018}
}

@article{liska2022streamingqa,
  title={StreamingQA: A Benchmark for Adaptation to New Knowledge over Time in Question Answering Models},
  author={Liska, Adam and Kocisky, Tomas and Gribovskaya, Elena and Terzi, Tayfun and Sezener, Eren and Agrawal, Devang and d'Autume, Cyprien de Masson and Scholtes, Tim and Manzagol, Mose and Young, Peter and others},
  journal={ICML},
  year={2022}
}

@article{kasai2022realtime,
  title={RealTime QA: What's the Answer Right Now?},
  author={Kasai, Jungo and Sakaguchi, Keisuke and Takahashi, Yoichi and Bras, Ronan Le and Asai, Akari and Yu, Xinyan and Radev, Dragomir and Smith, Noah A and Choi, Yejin and Inui, Kentaro},
  journal={NeurIPS},
  year={2022}
}

@article{vu2023freshllms,
  title={FreshLLMs: Refreshing Large Language Models with Search Engine Augmentation},
  author={Vu, Tu andIyyer, Mohit and Wang, Xuezhi and Constant, Noah and Wei, Jason and Wei, Jason and  others},
  journal={arXiv preprint arXiv:2310.03214},
  year={2023}
}

\clearpage

\newcommand{\AppSectionEntry}[3]{%
    \noindent
    \textbf{#1\hspace{1em}#2} \dotfill \textbf{#3}\par
    \vspace{0.8em}
}

\newcommand{\AppSubSectionEntry}[3]{%
    \noindent
    \hspace{2.5em} % 缩进
    \makebox[2.5em][l]{#1} % 编号宽度固定，保证对齐
    #2 \dotfill #3\par
    \vspace{0.4em}
}
% ==========================================
% 页面头部
% ==========================================
\noindent
{\Large \textbf{Appendix}} 
\hfill 

\vspace{1cm}

% ==========================================
% 目录标题
% ==========================================
\noindent
{\Large \textbf{Table of Contents}}

\vspace{1cm}

% ==========================================
% 目录内容
% ==========================================

% --- Section A: Prompts ---
\AppSectionEntry{A}{Prompt Templates}{10}

    \AppSubSectionEntry{A.1}{Workflow Generation Prompt}{10}
    
    \AppSubSectionEntry{A.2}{Workflow Repair Prompt}{10}

    \AppSubSectionEntry{A.3}{Model Generation Prompt}{11}
    
    \AppSubSectionEntry{A.4}{Evaluation Prompt}{11}

\vspace{1em} % 分隔不同大章节

% --- Section B: Trajectories ---
\AppSectionEntry{B}{Trajectories}{5}

    \AppSubSectionEntry{B.1}{Example of Case L1}{12}
    
    \AppSubSectionEntry{B.2}{Example of Case L2}{12}
    
    \AppSubSectionEntry{B.3}{Example of Case L3}{13}

\newpage

\definecolor{promptbg}{RGB}{250, 250, 250} 
\definecolor{promptframe}{RGB}{200, 200, 200} 
\definecolor{keywordcolor}{RGB}{0, 0, 128} 

\newtcolorbox{promptbox}[2][]{
    enhanced,
    breakable,
    colback=promptbg,
    colframe=promptframe,
    title={\textbf{#2}},
    coltitle=black,
    fonttitle=\large,
    attach boxed title to top left={yshift*=-\tcboxedtitleheight/2, xshift=5mm},
    boxed title style={
        colback=white,
        colframe=promptframe, 
        boxrule=0.5pt,     
        arc=2mm,           
        sharp corners=south, 
        left=2mm, right=2mm, top=1mm, bottom=1mm 
    },
    boxrule=0.5pt,
    arc=3mm,
    top=6mm,
    bottom=4mm, 
    left=4mm, 
    right=4mm,
    fontupper=\ttfamily\small, 
}

\appendix 

% --- Appendix A: Prompt Templates ---
\section{Prompt Templates}

This appendix provides the detailed prompt templates used in our experiments.

\subsection{Workflow Generation Prompt}
The executable code workflow prompt guides the agent to generate and iteratively validate an executable code workflow for time-sensitive questions, ensuring that the workflow can be successfully executed to retrieve answers from the target website based on the current time.
\begin{CJK*}{UTF8}{gbsn}

\begin{promptbox}{Workflow Generation Prompt Template}
\textbf{\textcolor{keywordcolor}{[System Prompt]}} \\

你是一个专业的代码生成和验证助手。你的任务是：
\begin{enumerate}[leftmargin=1.5em, topsep=2pt, itemsep=0pt]
    \item 基于给定的网站，生成和时间相关的问题，以及对应的获取问题答案的工作流代码。
    \item 使用\texttt{code\_interpreter}工具测试代码是否可执行。如果执行成功，则调用\texttt{check\_output\_content}工具进行最终检查；否则调用\texttt{think}工具思考问题原因，并修改代码，直到代码可执行为止。
    \item 使用\texttt{save\_generated\_item}工具保存最终可执行的问题和答案获取工作流代码。
    \item 关键结果需要\texttt{return}，\texttt{main}函数中使用\texttt{print}打印\texttt{return}的内容。
\end{enumerate}
\vspace{1em}
\textbf{要求：}
\begin{itemize}[leftmargin=1.5em, topsep=2pt, itemsep=0pt]
    \item 问题需要是实时性问题，跟时间相关。
    \item 代码里使用\texttt{from datetime import datetime; now = datetime.now()}获取时间，不可预设一个固定时间。
    \item 所有代码中不要包含任何\texttt{try except}，不要包含任何兜底策略。
    \item 工作流代码类似于爬虫代码，尽量仿照示例的代码样式，使用\texttt{playwright}和\texttt{markdownify}。
    \item 工作流代码必须经过测试确保可执行。
    \item 成功执行工作流代码后用\texttt{save\_generated\_item}保存。
\end{itemize}
\vspace{1em}
\textbf{示例：} \\
实时性问题：\{sample\_question\} \\
工作流代码：\{sample\_workflow\}

\vspace{1em}
\vspace{1em}

\textbf{\textcolor{keywordcolor}{[User Prompt]}} \\

\{URl and Example Question\}
\end{promptbox}

\end{CJK*} 

\subsection{Workflow Repair Prompt}
The workflow repair prompt guides the Repair Agent to diagnose execution failures of an existing Executable Code Workflow and to revise the workflow to restore stable execution and correct answer retrieval from the target website under the current time setting.

\begin{CJK*}{UTF8}{gbsn}
\begin{promptbox}{Workflow Repair Prompt Template}
\textbf{\textcolor{keywordcolor}{[System Prompt]}} \\

你是一个专业的代码修复和验证助手。你的任务是：基于给定的已有工作流代码，诊断代码存在的问题，找到原因并修复代码，确保代码能够在给定的网站上稳定获取实时性答案。

\vspace{1em}

\textbf{背景信息：}
\begin{itemize}[leftmargin=1.5em, topsep=2pt, itemsep=0pt]
    \item 你将收到一个之前可以正常使用的工作流代码，该工作流代码用于在指定网站上稳定获取实时性问题的答案。
    \item 该代码现在运行出现了问题（可能是网站结构变动、网址不可访问、元素定位失效等）导致无法获取正确答案。
    \item 你需要找到代码无法工作的原因并修复代码使其重新正常工作。
\end{itemize}

\vspace{1em}

\textbf{要求：}
\begin{itemize}[leftmargin=1.5em, topsep=2pt, itemsep=0pt]
    \item 代码修复后必须保持原有功能：回答的问题内容、最终输出格式都不能发生变化。
    \item 不允许使用硬编码数据作为兜底策略，必须确保代码能够获取真实的实时数据。
    \item 若网站结构发生变化，需要更新相应的元素定位和数据提取逻辑。
    \item 代码里使用\texttt{from datetime import datetime; now = datetime.now()}获取时间，不可预设一个固定时间。
    \item 所有代码中不要包含任何\texttt{try except}，不要包含任何兜底策略。
    \item 工作流代码类似于爬虫代码，参考给你的代码样式，使用 \texttt{playwright}和\texttt{markdownify}。
    \item 工作流代码必须经过测试确保可执行。
    \item 成功执行工作流代码后用\texttt{save\_generated\_item}保存。
\end{itemize}

\vspace{1em}

\textbf{工具使用技巧：}
\begin{enumerate}[leftmargin=1.5em, topsep=2pt, itemsep=0pt]
    \item 使用\texttt{check\_html\_content}工具可以帮助你检查网页中内容，或者确保网页中包含所需信息。
    \item 使用\texttt{think}工具深度分析问题原因和修复策略。
    \item 使用\texttt{code\_interpreter}工具测试代码是否可执行。如果执行成功，则调用\texttt{check\_output\_content}工具进行最终检查；否则调用\texttt{think}工具思考问题原因，并修改代码，直到代码可执行为止。
    \item关键结果需要\texttt{return}，\texttt{main}函数中使用\texttt{result}存储要\texttt{return}的内容，并使用\texttt{print}打印\texttt{result}。
    \item 使用\texttt{save\_generated\_item}工具保存通过验证的问题和答案获取工作流代码。
    \item 当你认为工作流的结果符合预期的时候，你就可以不再修改代码，而是保存。
\end{enumerate}

\vspace{1em}

\vspace{1em}

\textbf{\textcolor{keywordcolor}{[User Prompt]}} \\

存在问题的工作流: 

实时性问题: \{question\}

相应的工作流代码: \{workflow\}

\end{promptbox}

\end{CJK*} 

\subsection{Model Generation Prompt}
\label{app:model_generation_prompt}

The model generation prompt instructs the evaluated agent to explicitly anchor the current time, resolve relative temporal expressions before tool use, verify source freshness, and synthesize answers based on the most recent credible evidence.

\begin{promptbox}{The System Prompt for Model Generation}
You are a real-time deep research assistant. Your core function is to conduct rapid, accurate investigations into time-sensitive topics, current events, and dynamic data. You are explicitly designed to handle highly time-sensitive problems, where correctness depends heavily on the current date, recent events, or near-future conditions (e.g., weather forecasts, breaking news, policy updates, financial data, or daily headlines).

\vspace{0.5em}
\hrule
\vspace{0.5em}

\textbf{Real-Time Protocol:}
\begin{enumerate}[leftmargin=1.5em, topsep=2pt, itemsep=0pt]
    \item \textbf{Time Anchoring:} Immediately identify the ``Current Time'' provided at the end of this prompt. Use this as your absolute reference point.
    \item \textbf{Relative Time Resolution:} Before calling any tools, convert all relative time expressions in the user's query (e.g., ``yesterday'', ``next Friday'', ``last month'') into specific calendar dates (YYYY-MM-DD).
    \item \textbf{Source Verification:} Explicitly verify the publication date of any webpage or article you visit to ensure the information is not outdated.
    \item \textbf{Synthesis:} If sources conflict, prioritize the most recent credible update.
\end{enumerate}

\vspace{0.5em}
\textit{Current time: [Inserted Dynamically, e.g., 2026-01-15 17:45:27]}
\end{promptbox}

\subsection{Evaluation Prompt}
The evaluation prompt instructs the LLM-as-a-Judge to extract the final answer from the model response and determine whether it is semantically equivalent to the provided ground-truth answer, producing a binary correctness judgment.

\begin{CJK*}{UTF8}{gbsn}

\begin{promptbox}{Evaluation Prompt Template}
请根据下方给出的明确且无歧义的[correct\_answ\allowbreak er]，判断以下[response]对[question]的回答是否正确。

\vspace{1em}
[question]: \{question\}

\vspace{1em}
[response]: \{response\}

\vspace{1em}
你的判断必须符合以下格式与标准: 

\vspace{1em}
\textbf{\textcolor{keywordcolor}{extracted\_final\_answer}}: 从[response]中提取出的最终且确切的答案。如果在response中找不到明确的最终答案，请写"None"。

\vspace{1em}
\textbf{\textcolor{keywordcolor}{[correct\_answer]}}: \{correct\_answer\}

\vspace{1em}
\textbf{\textcolor{keywordcolor}{reasoning}}: 说明为什么extracted\_final\_answer与[co rrect\_answer]相比是正确或错误的，仅关注两者之间是否存在实质性的差异。不要讨论题目背景，不要尝试重新计算或解题，也不要为与[correct\_answer]不同的答案进行辩护，只需判断答案是否一致。

\vspace{1em}
\textbf{\textcolor{keywordcolor}{correct}}: 如果extracted\_final\_answer与给出的[corr ect\_answer]完全一致，或在数值题中处于允许的微小误差范围（四舍五入）内，则填"yes"；否则（存在任何不一致、歧义、非等价或答案错误）填"no"。
\end{promptbox}

\end{CJK*} 

\newpage
\section{Trajectories}

% ==========================================
% Trajectory 专用样式定义
% ==========================================

% 定义颜色
\definecolor{trajbg}{RGB}{255, 255, 255}      % 背景纯白
\definecolor{trajframe}{RGB}{100, 100, 100}   % 边框深灰
\definecolor{roleblue}{RGB}{0, 51, 102}       % 角色标签颜色 (深蓝)
\definecolor{actionbg}{RGB}{245, 245, 245}    % Action 背景 (浅灰)
\definecolor{obsbg}{RGB}{240, 255, 240}       % Observation 背景 (浅绿)

% 定义 Trajectory 大框
\newtcolorbox{trajectorybox}[2][]{
    enhanced,
    breakable,
    colback=trajbg,
    colframe=trajframe,
    title={\textbf{#2}}, % 标题 (例如 B.1 Example of Case L1)
    coltitle=white,
    fonttitle=\large,
    attach boxed title to top left={yshift*=-\tcboxedtitleheight/2, xshift=5mm},
    boxed title style={
        colback=trajframe, % 标题背景深灰
        colframe=trajframe,
        boxrule=0.5pt,
        arc=2mm,
        left=2mm, right=2mm, top=1mm, bottom=1mm
    },
    boxrule=0.8pt,
    arc=3mm,
    top=6mm, bottom=4mm, left=4mm, right=4mm,
    fontupper=\ttfamily\small,
    #1
}

% 定义内部步骤的宏 (简化正文代码)

% 1. System Prompt / User Prompt
\newcommand{\stepRole}[2]{%
    \par\vspace{3mm}\noindent
    \textbf{\textcolor{roleblue}{[#1]}} \hfill \textit{#2} % 右侧显示时间
    \par\vspace{1mm}
}

% 2. Action (Tool Call)
\newcommand{\stepAction}[2]{%
    \par\vspace{2mm}\noindent
    \begin{tcolorbox}[
        colback=actionbg, colframe=actionbg,
        left=2mm, right=2mm, top=1mm, bottom=1mm,
        arc=1mm, boxrule=0pt
    ]
    \textbf{\textcolor{roleblue}{[Action]}} \\
    {\footnotesize #1}
    \end{tcolorbox}
}

% 3. Observation (Tool Output)
\newcommand{\stepObservation}[1]{%
    \par\vspace{1mm}\noindent
    \begin{tcolorbox}[
        colback=obsbg, colframe=obsbg,
        left=2mm, right=2mm, top=1mm, bottom=1mm,
        arc=1mm, boxrule=0pt
    ]
    \textbf{\textcolor{roleblue}{[Observation]}} \\
    {\footnotesize #1}
    \end{tcolorbox}
}

% 4. Final Answer
\newcommand{\stepFinalAnswer}[1]{%
    \par\vspace{4mm}\noindent
    \begin{tcolorbox}[
        colback=white, colframe=roleblue,
        title={\textbf{[Final Answer]}},
        coltitle=white, colbacktitle=roleblue,
        left=2mm, right=2mm, top=2mm, bottom=2mm,
        arc=2mm, boxrule=0.5pt
    ]
    #1
    \end{tcolorbox}
}

This section demonstrates the step-by-step execution process of our agent on different types of queries.

% B.1 Example of Case L1
\subsection{Example of Case L1}

\begin{trajectorybox}{B.1 Example of Case L1}
\begin{CJK*}{UTF8}{gbsn}

% System Prompt
\stepRole{System Prompt}{} 
\textit{[This part follows the \textbf{System Prompt for Model Generation} defined in Figure 4. ]}

% User Prompt
\stepRole{User Prompt}{} 
最近3天（不含今天）NBA比赛中得分最高的球队是哪支，得了多少分？

\stepRole{Agent Trajectory}{}

% Round 1
\stepAction{\textbf{Think:} 用户询问最近3天（不含今天）NBA比赛中得分最高的球队是哪支，得了多少分？
首先根据时间锚点：当前日期是2026-01-13 14:54:09
"最近3天（不含今天）"意味着：
- 不包含今天（2026-01-13）
- 包含过去的3天：2026-01-12、2026-01-11、2026-01-10
我需要查找这3天内所有NBA比赛的比分，找出单场得分最高的球队。

\textbf{Search:} "query": ["NBA January 10 2026 scores results", "NBA January 11 2026 scores results", "NBA January 12 2026 scores results", "NBA 2026-01-10 game results", "NBA 2026-01-11 game results", "NBA 2026-01-12 game results"]
}

\stepObservation{
\textbf{Search Results:}
1. [NBA results: January 10, 2026] ...
3. [NBA's Nightly Recap | January 10, 2026] Game Recap: Hornets 150, Jazz 95...
...
1. [NBA results: January 11, 2026] New Orleans Pelicans vs Orlando Magic (118–128)...
...
1. [NBA results: January 12, 2026] Kings 124-112 Lakers...
}

% Final Answer
\stepFinalAnswer{
根据2026年1月10日至12日，最近3天，不含今天的N BA比赛数据：

\textbf{得分最高的球队：夏洛特黄蜂队}

\textbf{得分：150分}

\textbf{比赛详情：}
\begin{itemize}[leftmargin=1.5em]
    \item \textbf{日期}：2026年1月10日
    \item \textbf{对阵}：夏洛特黄蜂  vs 犹他爵士
    \item \textbf{比分}：黄蜂 150 - 95 爵士
    \item \textbf{备注}：这是NBA历史上单场第二高得分纪录
\end{itemize}

这三天其他高分球队包括：
\begin{itemize}[leftmargin=1.5em]
    \item 1月11日：奥兰多魔术 128分 vs 新奥尔良鹈鹕
    \item 1月11日：亚特兰大老鹰 124分 vs 金州勇士
    \item 1月12日：萨克拉门托国王 124分 vs 洛杉矶湖人
\end{itemize}

数据来源：NBA官方网站、ESPN、Basketball- Reference等权威体育媒体
}

\end{CJK*} 
\end{trajectorybox}

% B.2 Example of Case L2
\subsection{Example of Case L2}

\begin{trajectorybox}{B.2 Example of Case L2}
\begin{CJK*}{UTF8}{gbsn}

% System Prompt
\stepRole{System Prompt}{} 
\textit{[This part follows the \textbf{System Prompt for Model Generation} defined in Figure 4. ]}

% User Prompt
\stepRole{User Prompt}{} 
某个联赛2025年10月28日的比赛中，波特兰开拓者以122:108战胜洛杉矶湖人，最近3天（不含今天）该联赛比赛中得分最高的球队是哪支，得了多少分？

\stepRole{Agent Trajectory}{}

% Round 1
\stepAction{\textbf{Think:} 用户询问：某个联赛2025年10月28日的比赛中，波特兰开拓者以122:108战胜洛杉矶湖人，最近3天（不含今天）该联赛比赛中得分最高的球队是哪支，得了多少分？
首先，我需要理解这个查询：
1. 有一个联赛在2025年10月28日有比赛
2. 那天波特兰开拓者122:108战胜洛杉矶湖人
3. 用户问的是"最近3天（不含今天）"该联赛中得分最高的球队
当前日期是2026年1月13日。所以"最近3天（不含今天）"指的是：
- 不含2026年1月13日
- 应该是2026年1月12日、1月11日、1月10日这三天
我初步判断这可能是NBA联赛的比赛。接下来需要查找具体信息来确认。

\textbf{Search:} "query": ["NBA scores January 12 2026", "NBA scores January 11 2026", "NBA scores January 10 2026", "NBA highest scoring team last 3 days"]
}

\stepObservation{
\textbf{Search Results:}
1. [Kings 124-112 Lakers (Jan 12, 2026)] ...
4. [NBAs Nightly Recap | January 12, 2026] Game Recap: Magic 128, Pelicans 118...
...
1. [Atlanta Hawks vs Golden State Warriors Jan 11, 2026] ...
...
1. [San Antonio Spurs vs Boston Celtics Jan 10, 2026] ...
3. [Mavericks vs Bulls, January 10, 2026] DAL (107) vs CHI (125)...
}

% Round 2
\stepAction{\textbf{Think:} 让我整理一下最近3天（2026年1月10日、11日、12日）的NBA比赛情况：
从搜索结果中，我看到：
**1月12日比赛：**
- 国王 124-112 湖人（国王得124分）
- 魔术 128-118 鹈鹕（魔术得128分）
**1月11日比赛：**
- 老鹰 vs 勇士（结果未显示完整）
- 国王 111-98 火箭（国王得111分）
**1月10日比赛：**
- 马刺 100-95 凯尔特人（马刺得100分）
- 公牛 125-107 独行侠（公牛得125分）
目前看来，最近3天得分最高的球队是奥兰多魔术，在1月12日对阵鹈鹕的比赛中得到128分。
但我需要确认1月11日是否有球队得分超过128分。让我搜索更详细的1月11日比赛结果。

\textbf{Search:} "query": ["January 11 2026 NBA scores complete results", "NBA January 11 2026 all game scores"]
}

\stepObservation{
\textbf{Search Results:}
3. [NBA results: January 11, 2026] ... New Orleans Pelicans vs Orlando Magic (118–128) · Brooklyn Nets vs Memphis Grizzlies (98–103)...
7. [Hawks 124-111 Warriors (Jan 11, 2026)] Atlanta Hawks @ Golden State Warriors · Alexander-Walker's 24 points lead 6 Hawks in double figures as Atlanta beats Warriors 124-111.
}

% Final Answer
\stepFinalAnswer{
根据最近3天（2026年1月10日、11日、12日）的NBA比赛数据，得分最高的球队是**奥兰多魔术**，他们在2026年1月12日的比赛中得到**128分**（以128:118战胜新奥尔良鹈鹕）。

具体这三天的高分球队情况如下：

\textbf{2026年1月12日}：
\begin{itemize}[leftmargin=1.5em]
    \item 奥兰多魔术 128分（最高）
    \item 克利夫兰骑士 123分
    \item 犹他爵士 123分
\end{itemize}

\textbf{2026年1月11日}：
\begin{itemize}[leftmargin=1.5em]
    \item 金州勇士 124分（当天最高）
    \item 萨克拉门托国王 111分
\end{itemize}

\textbf{2026年1月10日}：
\begin{itemize}[leftmargin=1.5em]
    \item 芝加哥公牛 125分（当天最高）
    \item 印第安纳步行者 123分
\end{itemize}

因此，奥兰多魔术在1月12日得到的128分是这三天所有比赛中的最高分。
}

\end{CJK*} 
\end{trajectorybox}

% B.3 Example of Case L3
\subsection{Example of Case L3}

\begin{trajectorybox}{B.3 Example of Case L3}
\begin{CJK*}{UTF8}{gbsn}

% System Prompt
\stepRole{System Prompt}{} 
\textit{[This part follows the \textbf{System Prompt for Model Generation} defined in Figure 4. ]}

% User Prompt
\stepRole{User Prompt}{} 
某个联赛2025年10月28日的比赛中，波特兰开拓者以122:108战胜了Dalton Knecht当时所在的球队，最近3天（不含今天）该联赛比赛中得分最高的球队是哪支，得了多少分？

\stepRole{Agent Trajectory}{}

% Round 1
\stepAction{\textbf{Think:} 用户提出了一个问题：在某个联赛2025年10月28日的比赛中，波特兰开拓者以122:108战胜了Dalton Knecht当时所在的球队... (省略部分思考) ...我们需要先确定这场比赛所属的联赛（很可能是NBA），然后查看2025年10月28日之后的三天内（即2025年10月29日、30日和31日）该联赛的所有比赛，找出其中单场得分最高的球队及其得分。
需要注意的是：“最近3天（不含今天）”指的是相对于当前时间（2026年1月13日）的过去三天吗？... (省略部分思考) ...解读如下：该问题描述的是2025年10月28日的一场联赛比赛，然后询问“在过去三天内（不包括今天），该联赛中哪支球队在单场比赛中得分最高，具体得分是多少？” 因此，问题的重点是基于当前日期（2026年1月13日）往前推三天，即2026年1月10日至12日。
首先，我们需要确认Dalton Knecht在2025年10月28日时所属的球队。他是洛杉矶湖人队的新秀吗？实际上，我们需要核实这一点。让我们先搜索Dalton Knecht的相关信息。

\textbf{Search:} "query": ["Dalton Knecht October 2025 team", "Dalton Knecht 2025 Portland Trail Blazers 122-108", "Dalton Knecht NBA 2025", "Dalton Knecht traded 2025"]
}

\stepObservation{
\textbf{Search Results:}
1. [Dalton Knecht 2025-26 Stats per Game - NBA] ...
2. [Dalton Knecht, Los Angeles Lakers, SF - 2025-26 Game Log] ... Oct 27, 2025, vs POR. L 122-108.
...
}

% Round 2
\stepAction{\textbf{Think:} 搜索结果显示，2025年10月27日（或10月28日？）波特兰开拓者队与洛杉矶湖人队之间的一场比赛中，提到了Dalton Knecht。结果显示“波特兰开拓者队以122比108战胜湖人队（10月27日）”。... (省略部分思考) ...然而，关键信息是：该联赛为NBA，Dalton Knecht当时效力于洛杉矶湖人队。
现在回到问题：“最近3天（不含今天）该联赛比赛中得分最高的球队是哪支，得了多少分？” 意思是“在过去三天内（不包括今天），哪支球队在该联赛中取得了最高的单场得分，得分是多少？”
我们需要查看从今天（2026年1月13日）起往前推三天的比赛情况，即2026年1月10日、11日和12日（因为排除了今天）。需要在这三天中找出NBA所有比赛里单场得分最高的球队。

\textbf{Search:} "query": ["NBA scores January 10 2026", "NBA scores January 11 2026", "NBA scores January 12 2026", "NBA results Jan 10-12 2026"]
}

\stepObservation{
\textbf{Search Results:}
1. [San Antonio Spurs vs Boston Celtics Jan 10, 2026 Box ...] ...
...
6. [NBA's Nightly Recap | January 10, 2026] Game Recap: Hornets 150, Jazz 95...
...
}

% Round 3
\stepAction{\textbf{Think:} 我们有许多比赛结果。需要找出2026年1月10日、11日和12日这三天中，哪支球队在单场比赛中得分最高。
我们需要列出这些比赛中每支球队的得分，然后找出最高分。
可以通过搜索更详细的比分汇总，或逐一查看每场比赛的得分情况来实现。
让我们先获取2026年1月10日的比赛结果。... (省略部分思考) ...此外，可能还有更多比赛：例如独行侠 107-125 公牛？等等，有一条结果是“独行侠 vs 公牛，2026年1月10日”，比分为“DAL (107) vs CHI (125)”。这表明公牛队得了125分，是一个较高的得分。不过，我们需要确认该信息是否属实。

\textbf{Search:} "query": ["January 10 2026 NBA high score", "NBA January 10 2026 highest scoring team", "2026-01-10 NBA game scores list"]
}

\stepObservation{
\textbf{Search Results:}
6. [Post Game Thread - NBA: The Hornets defeat the Jazz on ...] Post Game Thread - NBA: The Hornets defeat the Jazz on Jan 10, 2026, the final score is 95-150.
...
}

% Round 4
\stepAction{\textbf{Think:} 搜索结果展示了1月10日的多场比赛。让我们提取每场比赛的得分。
从“NBA比赛结果：2026年1月10日——胜者与负者”页面中，我们看到“克利夫兰骑士队对阵明尼苏达森林狼队（146-134）”，这是一个很高的得分：骑士队得到146分。另外，“夏洛特黄蜂队对阵犹他爵士队（95-150）”？... (省略部分思考) ...因此，目前最高的是黄蜂队的150分。我们需要确认是否还有其他球队在1月10日至12日期间的任何比赛中得分超过150分。
现在来看1月11日的比赛。我们需要找出所有比赛中的最高得分。可能会有像“老鹰队124分，勇士队111分”这样的比赛（老鹰队得到124分）。或者其他队伍得到130分？让我们搜索1月11日的比赛结果。

\textbf{Search:} "query": ["NBA results January 11 2026 high score", "January 11 2026 NBA game scores list", "NBA January 11 2026 highest scoring team"]
}

\stepObservation{
\textbf{Search Results:}
1. [NBA results: January 11, 2026 – winners and losers] ... New Orleans Pelicans vs Orlando Magic (118–128)...
...
}

% Round 5
\stepAction{\textbf{Think:} 从1月11日的比赛结果来看：
- 新奥尔良鹈鹕队 vs 奥兰多魔术队（118–128）=> 魔术队得到128分。
- 布鲁克林篮网队 vs 孟菲斯灰熊队（98–103）=> 灰熊队得到103分。
... (省略部分思考) ...
我们需要找出1月10日至12日期间所有比赛中的最高得分。目前看来，1月10日黄蜂队得到的150分是最高的。但需要确认1月12日是否有球队得分超过150分。
搜索1月12日的比赛结果。

\textbf{Search:} "query": ["NBA results January 12 2026", "January 12 2026 NBA game scores", "NBA January 12 2026 high score"]
}

\stepObservation{
\textbf{Search Results:}
1. [NBA results: January 12, 2026 – winners and losers] ... Utah Jazz vs Cleveland Cavaliers (123-112)...
...
}

% Round 6
\stepAction{\textbf{Think:} 根据1月12日的比赛结果，部分比赛得分如下：
- 犹他爵士队对阵克利夫兰骑士队（123-112）=> 爵士队得到123分。
... (省略部分思考) ...
但我们需要找出1月10日至12日期间所有比赛中的最高得分。
从1月10日的比赛来看，黄蜂队得到150分（对犹他爵士队）。这似乎是最高的。但需要确认1月11日或12日是否有球队得分超过150分。让我们查看1月11日的比赛：奥兰多魔术队得到128分；老鹰队得到124分；雷霆队得到124分；其他球队可能更低。因此，没有比150分更高的。
1月12日：爵士队得到123分；国王队得到124分（也许？实际上国王队以124比112战胜湖人队）；还有其他队伍吗？快船队得到117分？没有超过150分的。
因此，最近三天（1月10日至12日）内单场得分最高的球队是夏洛特黄蜂队，他们在1月10日对阵犹他爵士队的比赛中得到了150分。... (省略部分思考) ...
现在需要回答问题：“最近3天（不含今天）该联赛比赛中得分最高的球队是哪支，得了多少分？”
因此答案为：夏洛特黄蜂队，150分。
但我们应再次确认是否有其他球队在1月11日或12日的比赛中得到超过150分。例如，1月11日是否有球队得到152分？可能性不大。不过我们仍需全面考虑：可以搜索“2026年1月11日 NBA 球队得分150”。

\textbf{Search:} "query": ["January 11 2026 NBA 150 points", "January 12 2026 NBA 150 points", "NBA 150 points January 2026"]
}

\stepObservation{
\textbf{Search Results:}
1. [Hornets 150-95 Jazz (11 Jan, 2026) Game Recap] ...
2. [Game Recap: Hornets 150, Jazz 95] ... Jan 11, 2026...
...
}

% Round 7
\stepAction{\textbf{Think:} 搜索结果显示，黄蜂队以150比95战胜爵士队的比赛发生在1月10日（部分来源显示为1月11日？）。实际上，多个来源显示该场比赛的日期为1月10日。ESPN的链接显示“黄蜂队150-95爵士队（1月10日，2026年）”。因此，该场比赛确实发生在1月10日。
现在我们需要确认最近三天（1月10日、11日、12日）内的最高得分是否为150分。1月11日是否还有球队得分超过150分？可能性不大。但我们需要再次确认是否有球队在1月11日或12日的比赛中单场得分达到或超过150分。例如，是否有一场比赛公牛队以152比150战胜老鹰队？那将是一场非常高得分的比赛。搜索结果显示“公牛队以152-150险胜老鹰队，创下NBA本赛季单场得分最高纪录”。这是一场得分极高的比赛。该条目出现在对“NBA 2026年1月150分”的搜索结果中。它可能指的是1月12日的一场比赛？让我们打开该链接查看。

\textbf{Visit:} "url": "https://slamonline.com/nba/bull s-edge-hawks-152-150-in-nba-s-highest-scoring -game-this-season/"
}

\stepObservation{
\textbf{Error:} The webpage content is inaccessible due to a 403 Forbidden error and CAPTCHA protection...
}

% Round 8
\stepAction{\textbf{Think:} 由于验证码限制，我们无法访问该页面。让我们尝试搜索“Bulls 152 Hawks 150 2026”以获取更多信息。

\textbf{Search:} "query": ["Bulls 152 Hawks 150 2026", "Chicago Bulls 152 Atlanta Hawks 150 January 2026", "Bulls edge Hawks 152-150 2026"]
}

\stepObservation{
\textbf{Search Results:}
1. [Bulls 152-150 Hawks (Dec 21, 2025) Final Score] ...
...
}

% Final Answer
\stepFinalAnswer{
根据查询结果，最近3天（2026年1月10日、11日、12日）NBA比赛中得分最高的球队是**夏洛特黄蜂队**，他们在2026年1月10日的比赛中以**150-95**大胜犹他爵士队。

**关键信息：**
\begin{itemize}[leftmargin=1.5em]
    \item \textbf{球队}：夏洛特黄蜂 (Charlotte Hornets)
    \item \textbf{得分}：150分
    \item \textbf{日期}：2026年1月10日
    \item \textbf{对手}：犹他爵士 (Utah Jazz)
    \item \textbf{比分}：150-95
\end{itemize}

这是近三天NBA所有比赛中的最高得分，也是NBA历史上单场第二高得分纪录。相比之下：
\begin{itemize}[leftmargin=1.5em]
    \item 1月11日最高分：奥兰多魔术 128分
    \item 1月12日最高分：萨克拉门托国王 124分
\end{itemize}

因此，夏洛特黄蜂队的150分是近期三天内的最高得分表现。
}

\end{CJK*} 
\end{trajectorybox}

% B.1 Example of Case L1
\subsection{Example of Case L1}

\begin{trajectorybox}{B.1 Example of Case L1}
\begin{CJK*}{UTF8}{gbsn}
% System Prompt
\stepRole{System Prompt} 
\textit{[This part follows the \textbf{System Prompt for Model Generation} defined in Figure 4. ]}

% User Prompt
\stepRole{User Prompt}{} 
最新一期双色球的开奖号码是什么？

\stepRole{Agent Trajectory}{}
% Round 1
\stepAction{\textbf{Think:} 用户现在需要最新一期双色球的开奖号码，首先得确定最新的开奖时间。双色球一般每周二、周四、周日开奖，现在当前时间是2026-01-12 14:00... (省略部分思考) ...所以应该搜索“最新一期双色球开奖号码”，这样工具会返回最新的结果。

\textbf{Search:} "query": ["最新一期双色球开奖号码"]
}

\stepObservation{
\textbf{Search Results:}
1. [双色球开奖结果\_中彩网] 2026-01-06（一）, 050609212830...
2. [双色球基本走势图] 彩球双色球：上期预测中一、二等1047万...
3. [双色球] 2026004, 03 07 08 09 18 32, 10...
4. [开奖信息\_中彩网] 双色球, 2026005期, 2026-01-11（日）, 01202227303310...
5. [双色球第2026004期] 1月8日晚...
...
}

% Round 2
\stepAction{\textbf{Think:} 用户现在需要最新一期双色球的开奖号码，首先得确定最新的开奖时间。双色球一般每周二、周四、周日开奖，现在当前时间是2026-01-12 14:00... (省略部分思考) ...所以应该搜索“最新一期双色球开奖号码”，这样工具会返回最新的结果。

\textbf{Visit:} "url": ["https://www.zhcw.com/kjxx/"], "goal": "确认2026005期双色球的开奖号码..."
}

\stepObservation{
\textbf{Search Results:}
Error. The provided webpage content could not be accessed. Please check the URL or file format.
}

% Round 2
\stepAction{\textbf{Think:} 刚才访问中彩网的开奖信息页面失败了，那换一个可靠的来源... (省略部分思考) ...再试一个搜索结果里的链接，比如第1条的中彩网双色球开奖结果页面，调用visit工具访问这个URL。

\textbf{Visit:} "url": ["https://www.zhcw.com/kjxx/ssq /"], "goal": "获取最新一期双色球的开奖号码"
}

\stepObservation{
\textbf{Search Results:}
最新一期双色球为2026005期，开奖日期是2026年1月11日（星期日），开奖号码为红球01、20、22、27、30、33，蓝球10。该期总销售额387903486元...
}

% Final Answer
\stepFinalAnswer{
最新一期双色球为第2026005期，开奖日期是2026年1月11日（星期日），具体开奖号码如下：
\begin{itemize}[leftmargin=1.5em]
    \item \textbf{红球号码}：01、20、22、27、30、33
    \item \textbf{蓝球号码}：10
\end{itemize}
该期总销售额387,903,486元，全国共开出4注一等奖，单注奖金10,000,000元，奖池累计金额达2,821,524,165元。
}
\end{CJK*} 
\end{trajectorybox}

% B.2 Example of Case L2
\subsection{Example of Case L2}

\begin{trajectorybox}{B.2 Example of Case L2}
\begin{CJK*}{UTF8}{gbsn}

% System Prompt
\stepRole{System Prompt}{} 
\textit{[This part follows the \textbf{System Prompt for Model Generation} defined in Figure 4. ]}

% User Prompt
\stepRole{User Prompt}{} 
某一彩票在2025年10月21日至2025年10月25日之间的某一期的中奖情况是：共开出6注大奖，其中吉林1注（1111.3448万元）、浙江1注（852.1343万元）、广东2注（分别为877.3925万元和850.3343万元）、重庆1注（850.3343万元）及甘肃1注（1111.3448万元）。该彩票最新一期的开奖号码是什么？

\stepRole{Agent Trajectory}{}

% Round 1
\stepAction{\textbf{Think:} 用户现在需要解决的问题是：先找到2025年10月21日至25日之间某一期开出6注大奖... (省略部分思考) ...应该先搜索这个中奖信息，确定是哪种彩票。

\textbf{Search:} "query": ["2025年10月21日至25日 彩票6注大奖 吉林1注1111万 甘肃1注1111万"]
}

\stepObservation{
\textbf{Search Results:}
1. [FMDM 封面代码] ...2. [案例--正义智库] ...... (Results are irrelevant) ...
}

% Round 2
\stepAction{\textbf{Think:} 刚才的搜索结果都是不相关的，可能是关键词不够准确。需要调整搜索关键词... (省略部分思考) ...试试搜索“2025年10月 大乐透 6注头奖 吉林 甘肃”，或者“2025年10月 双色球 6注头奖 吉林1111万”。

\textbf{Search:} "query": ["2025年10月 大乐透 6注头奖 吉林 甘肃"]
}

\stepObservation{
\textbf{Search Results:}
1. [双色球头奖6注850万分落5 地奖池余额28.7亿] 新浪彩票 2025年10月23日... 本期一等奖开出6注，单注奖金850万分落5地，其中广东2注，吉林、浙江、重庆、甘肃各1注...
}

% Round 3
\stepAction{\textbf{Think:} 第一个结果是新浪彩票2025年10月23日的双色球新闻... (省略部分思考) ...用户的信息里有1111万，这里是850万... 需要确认是否是同一期。

\textbf{Visit:} "url": ["https://lotto.sina.cn/2025-10 -23/detail-infuwwwk7754409.d.html..."]
}

\stepObservation{
\textbf{Page Content:}
北京时间2025年10月23日，福彩双色球第25122期开奖... 本期一等奖开出6注，单注奖金850万分落5地，其中广东2注，吉林、浙江、重庆、甘肃各1注...
}

% Round 4
\stepAction{\textbf{Think:} 地区分布完全一致，但金额不同。可能是派奖期间？... (省略部分思考) ...再搜索“双色球第25122期 派奖”。

\textbf{Search:} "query": ["双色球第25122期 派奖"]
}

\stepObservation{
\textbf{Search Results:}
6. [广东省福利彩票发行中心] ... 12 亿大派奖！... 2025-10-23 第2025122 期...
9. [双色球25122期开奖] 一等奖：共6注，每注奖金8,503,343元...
}

% Round 5
\stepAction{\textbf{Think:} 确认了，用户说的那期就是双色球第2025122期... (省略部分思考) ...现在需要查双色球最新一期的开奖号码，当前时间是2026年1月12日...

\textbf{Search:} "query": ["双色球最新一期开奖号码 2026年1月"]
}

\stepObservation{
\textbf{Search Results:}
2. [双色球开奖结果第2026005 期头奖4注奖金1000万] 2026年01月11日星期日... 开奖号码：红球为01 20 22 27 30 33，篮球为10...
}

% Final Answer
\stepFinalAnswer{
你所描述的是福彩双色球第2025122期（2025年10月23日开奖）的中奖情况，该期全国共开出6注一等奖，地区分布与奖金详情如下：
\begin{itemize}[leftmargin=1.5em]
    \item 吉林1注、甘肃1注：单注奖金含派奖后为1111.3448万元
    \item 广东2注：分别为877.3925万元和850.3343万元
    \item 浙江1注：852.1343万元
    \item 重庆1注：850.3343万元
\end{itemize}

截至2026年1月12日，双色球最新一期为第2026005期（2026年1月11日开奖），开奖号码为：
\textbf{红球：01、20、22、27、30、33}
\textbf{蓝球：10}
}

\end{CJK*} 
\end{trajectorybox}

% B.3 Example of Case L3
\subsection{Example of Case L3}

\begin{trajectorybox}{B.3 Example of Case L3}
\begin{CJK*}{UTF8}{gbsn}

% System Prompt
\stepRole{System Prompt}{} 
\textit{[This part follows the \textbf{System Prompt for Model Generation} defined in Figure 4. ]}

% User Prompt
\stepRole{User Prompt}{} 
A省的人民政府于2025年09月01日发布了关于同意设立珲乌高速公路拉法山收费站收取通行费的批复，某一彩票在2025年10月21日至2025年10月25日之间的某一期的中奖情况是：共开出6注大奖，其中A省1注（1111.3448万元）、浙江1注（852.1343万元）、广东2注（分别为877.3925万元和850.3343万元）、重庆1注（850.3343万元）及甘肃1注（1111.3448万元）。该彩票最新一期的开奖号码是什么？

\stepRole{Agent Trajectory}{}

% Round 1
\stepAction{\textbf{Think:} 首先需要确定“A省”是哪个省份。根据题目描述，A省人民政府于2025年9月1日发布了关于设立“珲乌高速公路拉法山收费站”的批复。我需要搜索这个收费站位于哪个省份。确定省份后，再根据时间范围（2025年10月21-25日）和中奖情况（6注大奖，A省、浙江、广东、重庆、甘肃各有多少注）来确定是哪种彩票的哪一期。最后查询该彩票在当前时间（2026年1月12日）的最新一期开奖号码。

\textbf{Search:} "query": ["珲乌高速公路拉法山收费站 省份", "2025年10月21日至25日 彩票开奖 6注大奖"]
}

\stepObservation{
\textbf{Search Results:}
1. [吉林省人民政府关于同意设立珲乌高速公路拉法山收费站...] 按照《收费公路管理条例》有关规定，现批复如下...
...
}

% Round 2
\stepAction{\textbf{Think:} 搜索结果明确显示“吉林省人民政府关于同意设立珲乌高速公路拉法山收费站...”，所以A省是**吉林省**。
接下来需要找到2025年10月21日至25日期间，吉林1注、浙江1注、广东2注、重庆1注、甘肃1注一等奖的彩票期数。这看起来像是双色球或大乐透。

\textbf{Search:} "query": ["2025年10月 彩票6注一等奖 吉林 甘肃 1111万元", "双色球25122期 开奖日期"]
}

\stepObservation{
\textbf{Search Results:}
1. [双色球头奖6注850万分落5地...] 2025年10月23日... 本期一等奖开出6注，分落吉林1注、浙江1注、重庆1注、广东2注、甘肃1注...
}

% Round 3
\stepAction{\textbf{Think:} 搜索结果完美匹配！双色球第2025122期（2025年10月23日开奖）的中奖分布与题目描述完全一致（吉林1、浙江1、广东2、重庆1、甘肃1）。虽然题目提到的金额（1111万）与基本奖金（850万）不同，但这通常是因为派奖活动导致的（双色球当时正在进行12亿大派奖）。
现在确定了彩票是**双色球**。当前时间是2026年1月12日（周一）18:30。双色球每周二、四、日开奖。1月12日是周日，但开奖时间通常是21:15，所以今晚的开奖还没出来。最新的一期应该是上周四（1月9日）的第2026004期。

\textbf{Search:} "query": ["双色球2026004期 开奖号码 官方"]
}

\stepObservation{
\textbf{Search Results:}
1. [双色球第2026004期开奖结果...] 红球号码为03、07、08、09、18、32，蓝球号码为10...
}

% Final Answer
\stepFinalAnswer{
根据你描述的彩票开奖信息（2025年10月23日开奖的双色球第2025122期，共开出6注一等奖，分布于吉林、浙江、广东2注、重庆、甘肃），可以确定该彩票为**中国福利彩票双色球**。

截至2026年1月12日18:30（双色球每周日21:15开奖，当期开奖尚未进行），最新已开奖的一期为**第2026004期**，开奖日期为2026年1月8日，具体开奖号码如下：
\begin{itemize}[leftmargin=1.5em]
    \item \textbf{红球}：03、07、08、09、18、32
    \item \textbf{蓝球}：10
\end{itemize}
当期全国共开出4注一等奖，单注奖金870万元，分别花落浙江2注、湖南1注、新疆1注。
}

\end{CJK*} 
\end{trajectorybox}

\end{document}